\documentclass[usegraphicx,useAMS,usenatbib]{mn2e}
\newcommand{\beq}{\begin{equation}}
\newcommand{\beqa}{\begin{eqnarray}}
\newcommand{\eeq}{\end{equation}}
\newcommand{\eeqa}{\end{eqnarray}}
\newcommand{\etahat}{\hat{\eta}}

\renewcommand{\vec}[1]{\bmath #1}
\title[Large resistivity in numerical simulations ...]{Large resistivity in
numerical simulations of radially self-similar outflows}
\author[M. \v{C}emelji\'{c} et al.]{M. \v{C}emelji\'{c}$^1$\thanks{E-Mail: miki@tiara.sinica.edu.tw
(M\v{C})}, N. Vlahakis$^2$\footnotemark[0], K. Tsinganos$^2$\footnotemark[0]
\\
$^1$Academia Sinica Institute of Astronomy and Astrophysics \& Theoretical
Institute for Advanced Research in Astrophysics,\\
P.O. Box 23-141, Taipei 10617, Taiwan\\
$^2$IASA and Section of Astrophysics, Astronomy and Mechanics,
Department of Physics, University of Athens,\\
Panepistemiopolis 15784 Zografos, Athens, Greece
}
\begin{document}

\date{Received/Accepted}
\pagerange{\pageref{firstpage}--\pageref{lastpage}} \pubyear{2010}
\maketitle
\label{firstpage}

\begin{abstract}
We investigate the differences between an outflow in a highly-resistive
accretion disk corona, and the results with smaller or vanishing
resistivity. For the first time, we determine conditions at the base of
a two-dimensional radially self-similar outflow in the regime of very
large resistivity. We performed simulations using the {\sc pluto}
magnetohydrodynamics code, and found three modes of solutions. The
first mode, with small resistivity, is similar to the ideal-MHD
solutions. In the second mode, with larger resistivity, the geometry of
the magnetic field changes, with a ``bulge'' above the super-fast critical
surface. At even larger resistivities, the third mode of solutions sets
in, in which the magnetic field is no longer collimated, but is
pressed towards the disk. This third mode is also the final one: it
does not change with further increase of resistivity. These modes
describe topological change in a magnetic field above the accretion
disk because of the uniform, constant Ohmic resistivity.
\end{abstract}


\section{Introduction}
Outflows of plasma in the vicinity of objects from stellar to galactic
scales have been observed, and are an essential ingredient in theoretical
models. The energy and angular momentum transport in any such system will most
certainly be affected by outflows. Their impact on the interstellar
medium is huge, and is probably part of the mechanism for recycling
stellar matter into new stars.

Because of computational difficulties, numerical simulations
with an increasing number of physical parameters can
not go much beyond the self-similar approach of, for example,
\citet{BP82}. This approach has been systematized in \citet{VT98} and
\citet{V00} into a general scheme for self-similar outflows, with two
sets of exact MHD outflow models, radially and meridionally self-similar.

Simulations in \citet{gra} confirmed the stationarity of such a solution in
the complete physical domain of a radially self-similar corona of a disk
in the ideal-MHD approach. In \citet{C08}(hereafter C08), we reported
an extension of radially self-similar numerical simulations in the
resistive-MHD simulations using the {\sc nirvana} code by \citet{udo}.
There we investigated the regime of small resistivity, and reported the
existence of apparent {\em critical magnetic diffusivity},
where the departure of the outflow shape from the ideal-MHD case
stops following the trend found for small physical resistivities.
Because of the long run-times of such simulations with the non-parallel
{\sc nirvana} (v.2) code, we could not proceed to investigate the
large resistivity regime.

Before we can indulge in the modeling of the anomalous magnetic
diffusivity in simulations instead of using the uniform diffusivity,
the eventual existence of its critical value has to be resolved. This
will then set the upper limit in models of resistivity, which could
otherwise be difficult to prescribe in the ramifications of a
particular model.

We introduce our setup, now with the parallel {\sc pluto}
code. Next we check for the influence of numerical resistivity, and
then address our results in the regime of large resistivity, as
compared to the trend found in simulations with small resistivity. We
then discuss possible consequences of our results for astrophysical
simulations.

\section{Problem setup}
We use a similar set of initial conditions as in C08, but now using the
{\sc pluto} (v. 3.1.1) code with the parallel option \citep{mig}. Here we
give only a short review, for particulars we refer the reader to
\cite{gra} and C08.

In SI units, the equations we solve are:
\beqa
\frac{\partial \rho}{\partial t} + \nabla \cdot (\rho \vec{V})=0 \,,\\
\rho\left[ \frac{\partial\vec{V} }{\partial t}+ \left( \vec{V
}\cdot \nabla\right) \vec{V} \right] + \nabla p +
\rho\nabla \Phi
- \frac{ \nabla \times \vec{B}}{\mu_0} \times \vec{B} = 0 \label{mom2} \,, \\
\frac{\partial\vec{B} }{\partial t}- \nabla \times \left( \vec{V}
\times \vec{B}-\eta \nabla \times \vec{B} \right)= 0 \,, \label{faraday}\\
\rho \left[ {\frac{\partial e}{\partial t}}
+ \left(\vec{V} \cdot \nabla\right)e \right]
+ p(\nabla \cdot\vec{V} )
- \frac{\eta}{\mu_0} \left( \nabla \times \vec{B} \right)^2= 0 \,, \label{enn}\\
 \nabla \cdot \vec{B}=0 \,,
\eeqa
with $\rho$ and $p$ for the density and pressure, $\vec{V}$ and $\vec{B}$
for the velocity and magnetic field, and $\Phi=-{\cal GM}/r$ as the
gravitational potential of the central mass ${\cal M}$. The internal
energy (per unit mass) is $e=p/(\rho(\gamma-1))$, where $\gamma$ is the
effective polytropic index, set to 1.05 on our simulations. The
magnetic diffusivity $\eta$ in the SI system of units is related to the
electrical resistivity $\eta_{\mathrm r}=\mu_0\eta$, where $\mu_0$ is
the permeability of vacuum. In cgs units, $\mu_0=1$ and
$\eta_{\mathrm r}=\eta$, so that in the text we will often refer to
$\eta$ as a resistivity, except when it could produce a misunderstanding.

\subsection{Initial and boundary conditions}
\begin{figure}
\includegraphics[width=7.5cm,height=8cm]{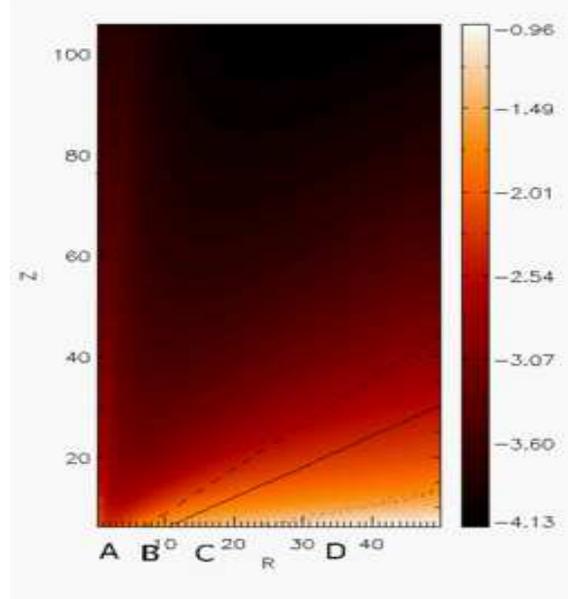}
\caption{Initial conditions in our simulations, with marked inner-Z
boundary regions of interest for the setup of boundary conditions. To
avoid reflection from the outer-R boundary, we actually set our
computational box three times larger in R-direction, and then
analyze results only in the $R\times Z=(128\times 256)$ grid cells
$=([0,50]\times [6,106])R_0$ portion of the domain. If not stated
otherwise, all the results in this paper are shown at such resolution.
In all the plots, the density is shown in a logarithmic color scale, and
the three critical surfaces are plotted in dashed, solid and dotted
lines, for fast magnetosonic, Alfv\'{e}n and slow
magnetosonic waves, respectively. Labels A, B, C and D mark
portions of the inner-Z boundary where the flow is super-fast
magnetosonic, super-Alfv\'{e}nic, super-sonic and sub-sonic,
respectively.
}
\label{initbcs}
\end{figure}
\begin{figure*}
\hspace{-.2cm}\includegraphics[width=5.5cm]{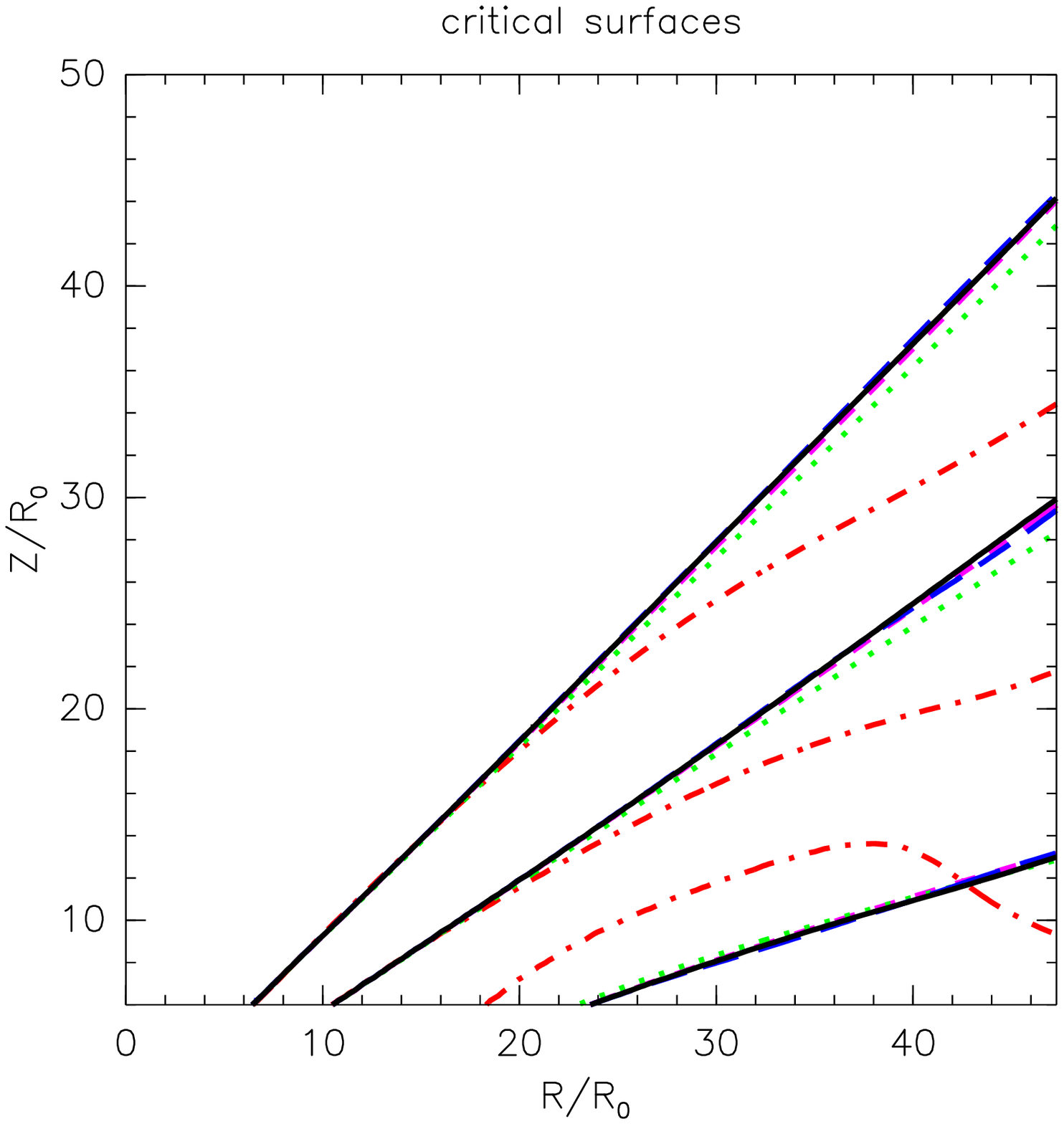}
\hspace{0cm}\includegraphics[width=5.5cm]{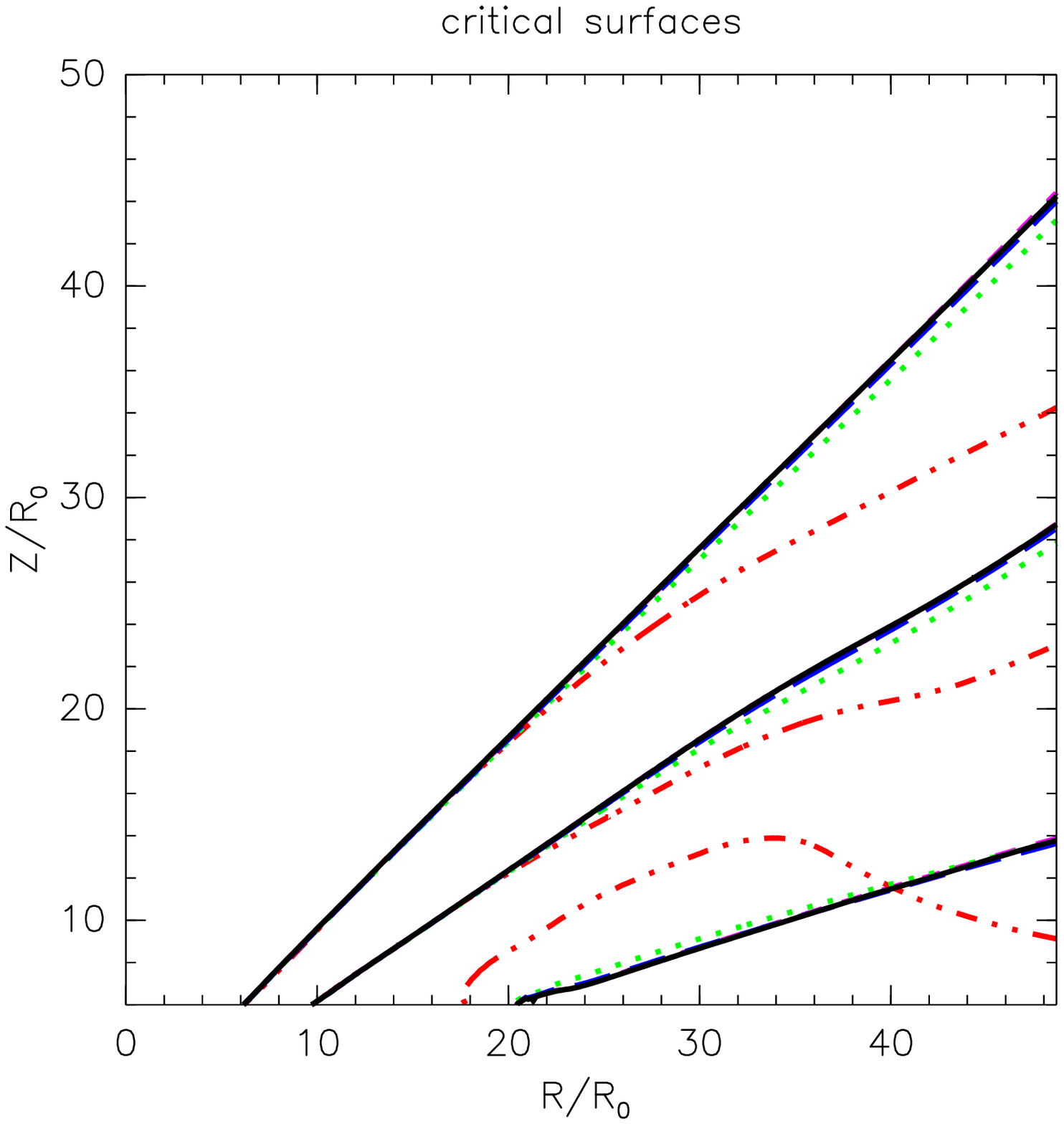} 
\hspace{0cm}\includegraphics[width=5.5cm]{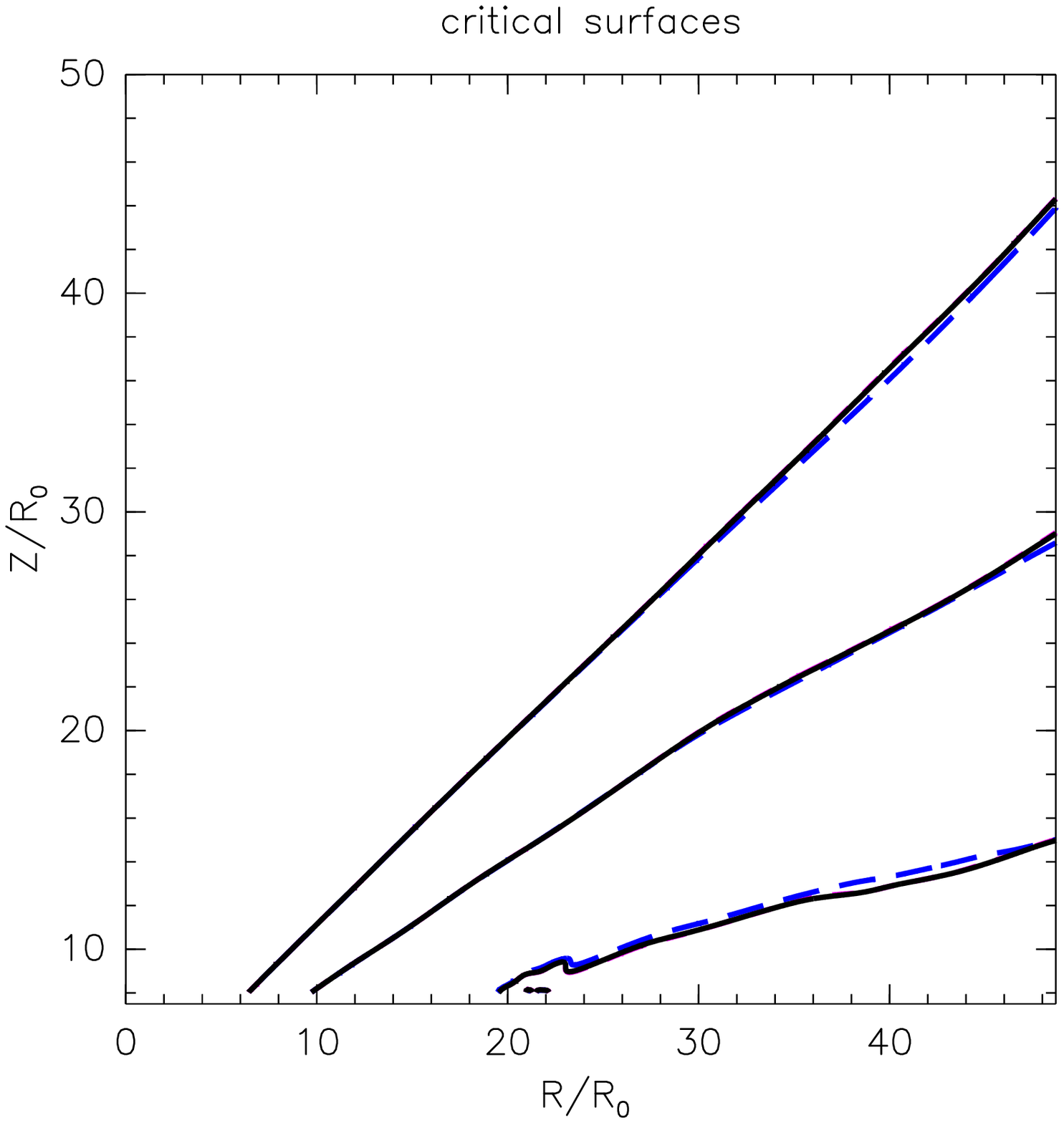} 
\caption{To estimate the numerical resistivity, we compare positions of
the critical surfaces in $R\times Z=(64\times 128)$ grid cells,
$R\times Z=(128\times 256)$ grid cells and $R\times Z=(384\times
768)$ grid cells in a $(50\times [6,106])R_0$ part of the computational
box in the {\em Left}, {\em Middle} and {\em Right} panels,
respectively. Shown is only half of the computational box in the
Z-direction, containing the critical surfaces. We show fast
magnetosonic, Alfv\'{e}n and slow magnetosonic critical surfaces
positioned from higher to lower positions in the box, respectively.
Solid (black), short-dashed (magenta), long-dashed (blue), dotted
(green) and dot-dot-dot-dashed (red) lines show the
quasi-stationary states (all at T=150) for
$\eta=0.0003, 0.003, 0.03, 0.15$ and 1.5, respectively. In the
rightmost panel we show results only for $\eta < 0.1$. From
inspection of these results it follows that the level of numerical
resistivity in $R\times Z=(64\times 128)$ and $(128\times 256)$
grid cells is about 0.1, and in the $R\times Z=(384\times 768)$
grid cells the numerical resistivity is about 0.01.
}
\label{numres}
\end{figure*}
\begin{figure*}
\hspace{0cm}\includegraphics[width=5.5cm,height=7.5cm]{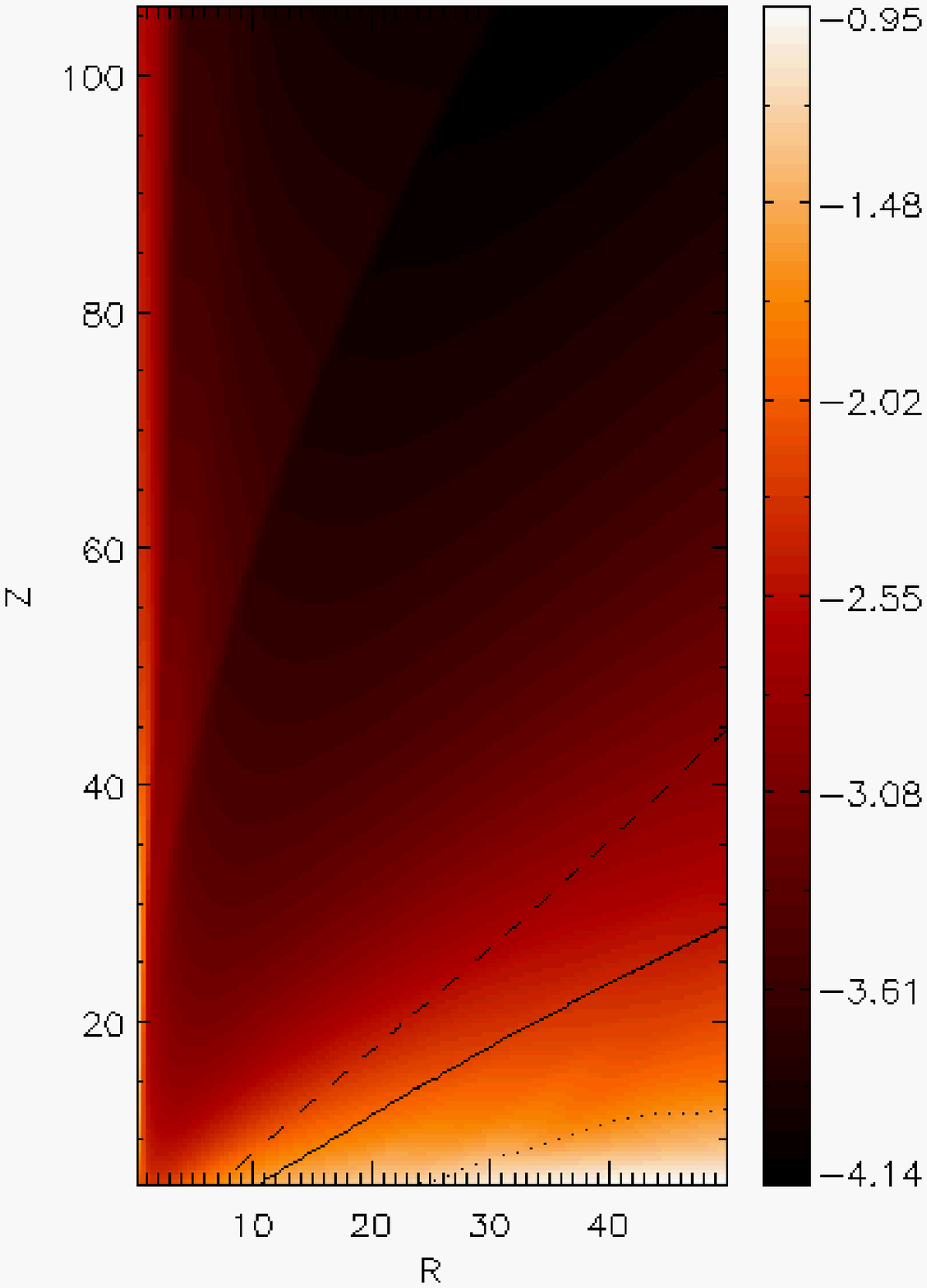} 
\includegraphics[width=5.5cm,height=7.5cm]{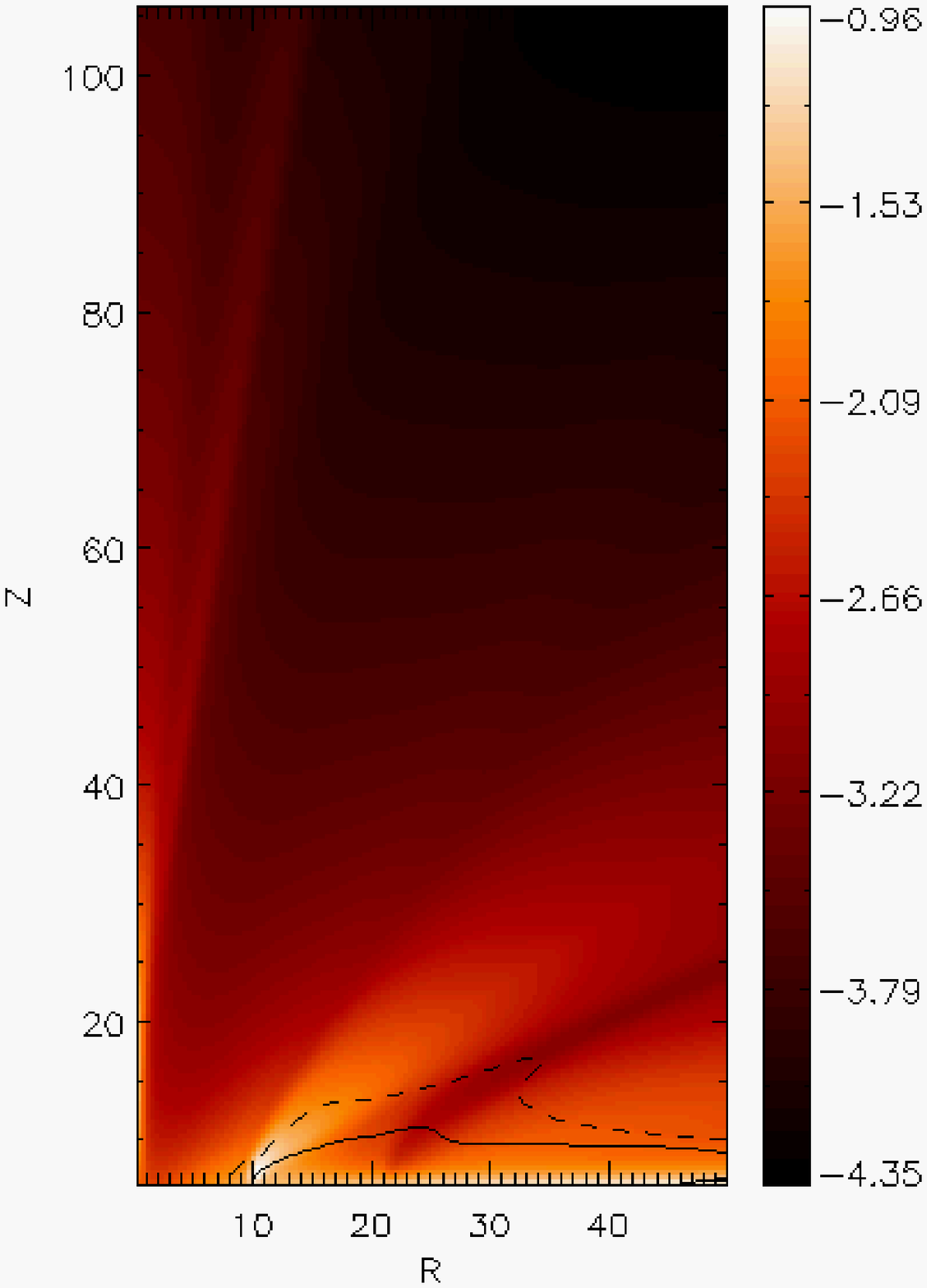}
\includegraphics[width=5.5cm,height=7.5cm]{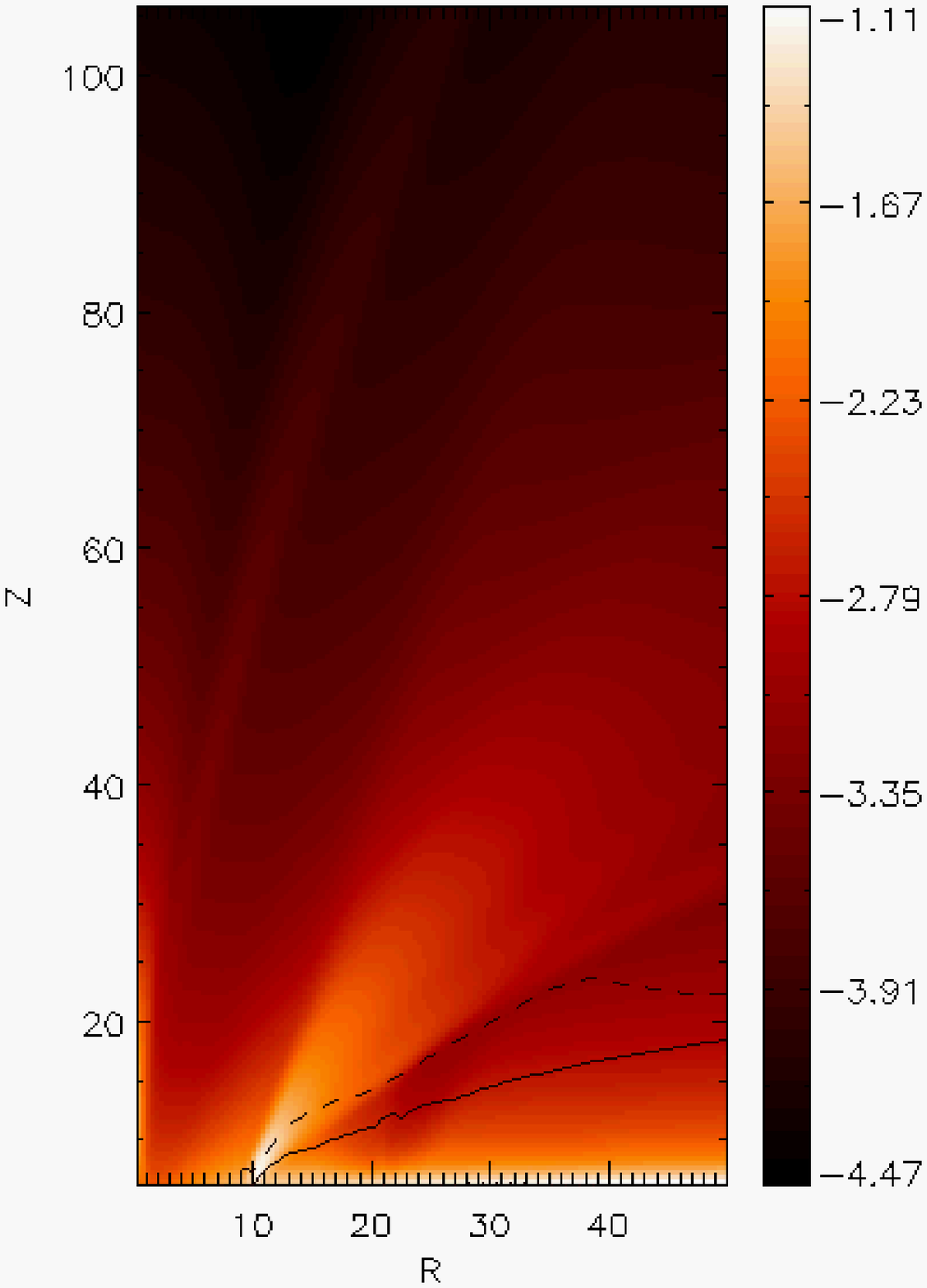} 
\hspace{-.2cm}\includegraphics[width=5.cm,height=6.5cm]{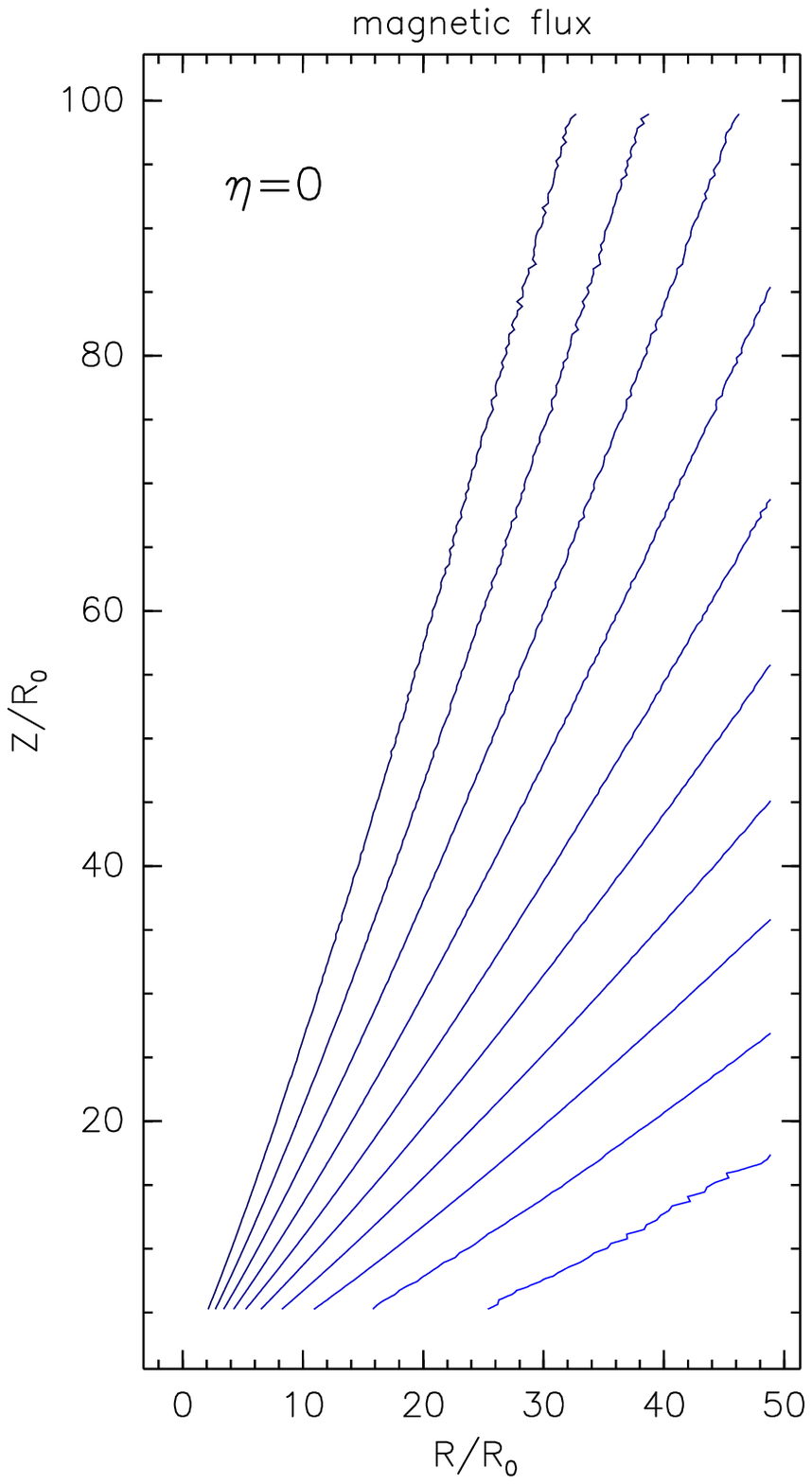}
\hspace{0cm}\includegraphics[width=5.cm,height=6.5cm]{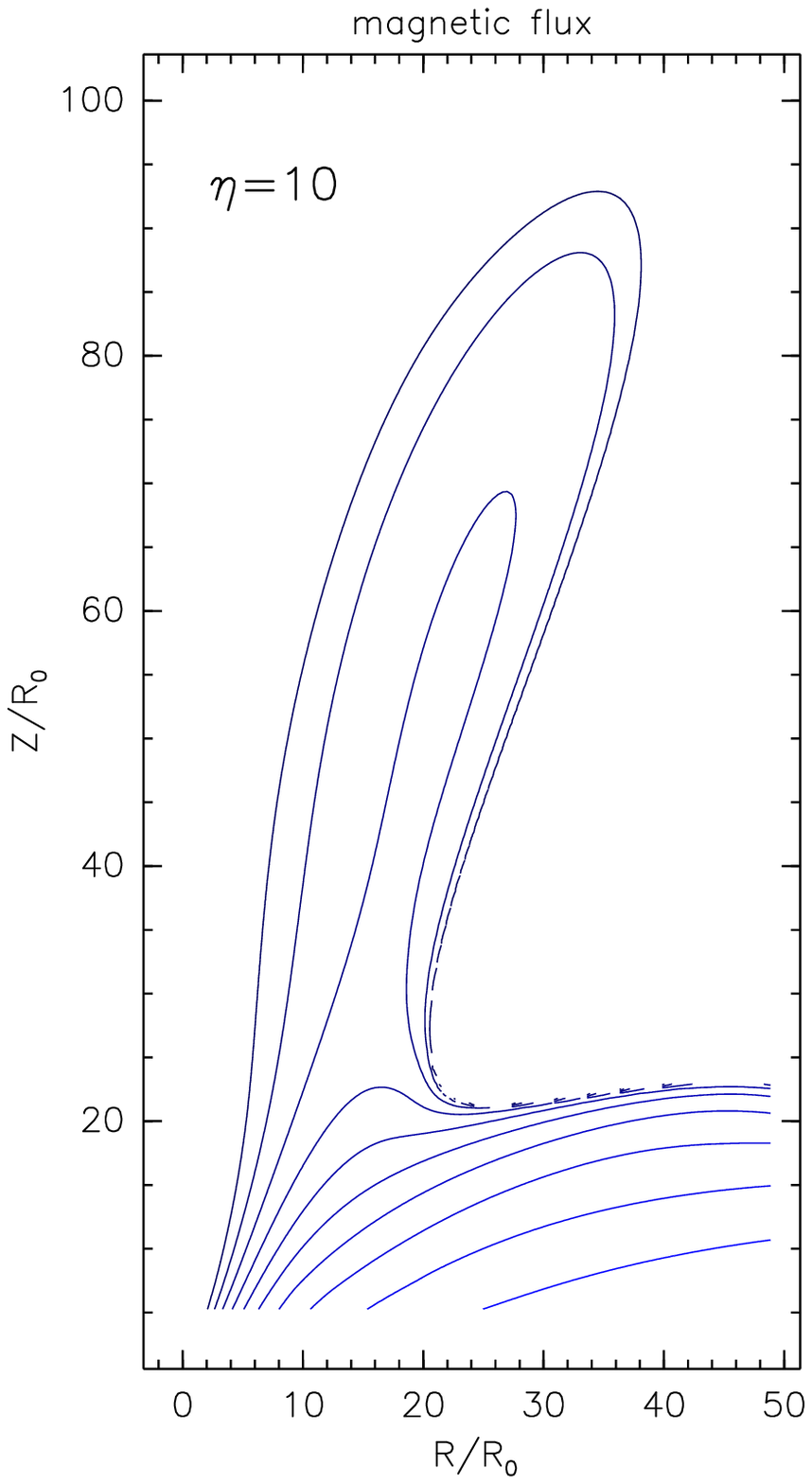} 
\hspace{0cm}\includegraphics[width=5.cm,height=6.5cm]{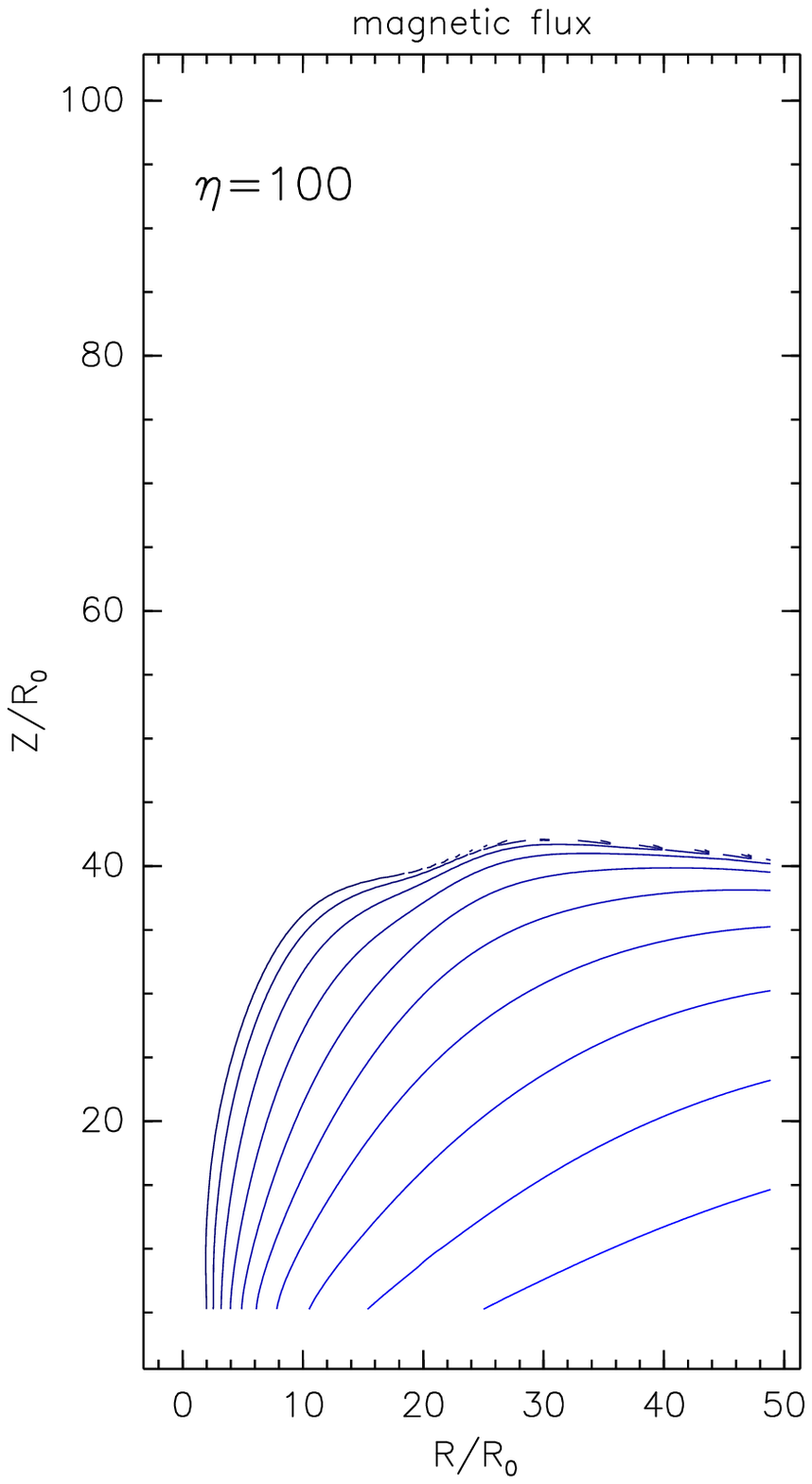}
\caption{Illustration of the effect of resistivity on the density in
the outflow. {\em Top panels}: In the {\em Left} panel are shown the
quasi-stationary state solutions in the ideal-MHD case, and in the
{\em Middle} and {\em Right} panels are shown solutions with large
and very large resistivity, $\eta=$ 10 and 100. Solutions with small
resistivity are very similar to the ideal-MHD solution.
Definitions of colors and lines are the same as in Figure \ref{initbcs}.
The obtained outflow for large and very large resistivity is
supersonic in the entire domain, so that the slow magnetosonic surface
is not present in those cases. {\em Bottom panels}: Magnetic flux
isocontour lines with the above resistivities, $\eta=0,10,100$
in the {\em Left}, {\em Middle} and {\em Right} panels, respectively.
Contours are parallel to the poloidal magnetic field lines.
With a small $\eta$, magnetic flux 
isocontour lines are of the same geometry as for the ideal-MHD case.
We obtain three different geometries of solutions for small, large and
very large resistivity. The topology of the magnetic field lines for
corresponding values of resistivity is always similar to one of the
three modes shown.
}
\label{densres}
\end{figure*}
We used the self-similar solution of \cite{V00} as the initial and
boundary conditions in the simulations. For steady-state, axisymmetric and
radially self-similar solutions we can write, in a form with mixed spherical
($r$, $\theta$, $\phi$) and cylindrical ($Z=r\cos\theta$, $R=r\sin\theta$,
$\phi$) coordinates:
\beqa
\frac{\rho}{\rho_0}=\alpha^{x-3/2}\frac{1}{M^2}\,,
\frac{p}{p_0}=\alpha^{x-2}\frac{1}{M^{2\gamma}}\,,\\
\frac{\vec{B}_p}{B_0}=-\alpha^{\frac{x}{2}-1}\frac{1}{G^2}\frac{
\sin\theta}{\cos(\psi+\theta)}
\left(\cos\psi \hat{R}+\sin\psi \hat{Z}\right) \,,
\label{bpss}\\
\frac{\vec{V}_p}{V_0}=-\alpha^{-1/4}\frac{M^2}{G^2}\frac{
\sin\theta}{\cos(\psi+\theta)}
\left(\cos\psi \hat{R}+\sin\psi \hat{Z}\right) \,, \\
\frac{{B}_\phi}{B_0}=-\lambda\alpha^{\frac{x}{2}-1}\frac{1-G^2}{G(1-M^2)}\,,
\frac{{V}_\phi}{{V}_0}=\lambda\alpha^{-\frac{1}{4}}\frac{G^2-M^2}{G(1-M^2)} \,,
\label{vphiss}
\eeqa
where $\displaystyle \alpha=\frac{R^2}{R_0^2 G^2}$,
$(M\,, G\,, \psi)$ are functions of $\theta$, and
\beq
\displaystyle{
V_0=\frac{1}{\kappa} \sqrt{\frac{ \cal G M }{R_0}} \,, \quad
\rho_0=\frac{B_0^2}{\mu_0 V_0^2} \,, \quad
p_0=\mu\frac{B_0^2}{2\mu_0} \,.
}
\label{norms}
\eeq
The index $p$ denotes poloidal, and the index $\phi$ toroidal components.

The density and pressure are related by a polytropic relation
$p=Q(\alpha)\rho^\gamma$, with the entropy function $Q(\alpha)$ constant
along a flow line, but differing from one flow line to the other. This
relation is the general steady state solution of Eq.~(\ref{enn}) in the
ideal-MHD case (without the last term) for the equation of state used here,
$p/rho=e/(\gamma-1)$. This procedure reduces the system of ideal-MHD
equations to three first-order ordinary differential equations with respect
to the functions $(G, M, \Psi)$. A particular solution is given by the set
of formal solution parameters
$(x\,, \lambda^2, \mu\,, \kappa\,, \gamma)=(0.75\,, 136.9\,, 2.99\,, 2\,, 1.05)$
and a prescription for the solution functions $(G, M, \Psi)$ is calculated
numerically by solving the first-order ordinary differential equations.
After ensuring that the flow crosses the three singular MHD surfaces, by
applying the appropriate regularity conditions \citep{V00}, the solution
free parameters are chosen by following the \cite{BP82} choice, for easier
comparison.

The magnetic diffusivity $\eta$ we included in the simulations as
a constant in the whole domain and normalized as
\beq
\eta = \etahat \ V_0 R_0=\etahat \sqrt{{\cal GM}R_0}/\kappa\,,
\label{magdif}
\eeq
with dimensionless $\etahat$. 

The self-similar solution breaks down near the rotation axis and the
analytical solution of \cite{V00} is not provided for $\theta$ smaller
than 0.025 rad, measured from the axis. To perform numerical simulations
in a computational box with the symmetry axis included, we modified the
analytical solution\footnote{In this modified solution, results for
$\theta$ smaller than 0.025 rad, measured from the axis, are also missing.
We linearly extrapolated it for the tabulated functions $G$, $M$ and
$\psi$. Modification of the functions G and M also means that the
pressure/energy is modified near the axis.}.

To maintain the divergence-free magnetic field as given by Eq.~(\ref{bpss})
with the extrapolated functions $G$ and $\psi$, the initial magnetic field
has to be modified. This is why, instead of using Eq.~(\ref{bpss}), we
compute the $B_Z$ component from the $\hat{Z}$
component of the self-similar expression
\beq
\vec{B}_p=\frac{B_0 R_0^2}{x} \nabla \times
\left( \alpha^{x/2}\frac{\hat{\phi}}{R} \right) \,,
\eeq
and the component $B_R$ from $\nabla\cdot \vec B_{\mathrm p}=0$, with
boundary condition $B_R(R=0)=0$. This procedure also
requires the modification of the poloidal velocity, to obtain
$\vec{V}_p\parallel \vec{B}_p$ of the steady initially ideal-MHD flow.
We compute the new direction of the poloidal initial velocity,
maintaining the speed. For the cases with included physical
resistivity, such a constraint will not be valid after the initial
moment, which we set by ideal-MHD requirements.

Boundary conditions are the symmetry along the rotation axis, which is
the inner, R$_{\rm min}$ boundary, and outflow conditions along the
outer $R$ and $Z$ boundaries. At the Z$_{\rm min}$ boundary we need to
constrain only one quantity \footnote{In C08 we fixed the boundary
values for six physical quantities, to maintain the constant magnetic
flux along the Z$_{\rm min}$ boundary, while in \cite{gra} the
boundaries were numerically over-specified, with 7 of them defined by
the initial values.}. A detailed description of the
boundary conditions is given
in \citet{M08}. In Figure \ref{initbcs} we show regions A, B, C, D where
$V_Z>V_{Z,fast}$, $V_{Z,fast}>V_Z>V_{Z,Alf}$, $V_{Z,Alf}>V_Z>V_{Z,slow}$
and $V_Z<V_{Z,slow}$, respectively. Of eight boundary conditions for
eight physical quantities (density, pressure, and three components for
each of velocity ${\mathrm V}_{\mathrm R}, {\mathrm V}_{\varphi},
{\mathrm V}_{\mathrm Z}$ and
magnetic field ${\mathrm B}_{\mathrm R}, {\mathrm B}_{\varphi},
{\mathrm B}_{\mathrm Z}$), one for magnetic field is
determined by the $\nabla\cdot\vec{B}=\nabla\cdot\vec{B_{\mathrm p}}=0$.
Of the remaining seven, three are determined from the computational
box, by linear extrapolation to the
ghost zone after crossing the three magnetosonic critical surfaces.
This is because the number of boundary conditions is reduced by one for
each critical surface which is crossed downstream, since the
corresponding magnetosonic waves can not propagate outwards along the
flow from those surfaces. We tried various combinations of quantities which
are extrapolated and which are not. The one in which we could obtain
simulations with the same setup for all the $\eta$, which enables
a good comparison between the results, is the one with the R-component of
velocity and the Z-component of the magnetic field extrapolated in
portion B of the boundary. In addition to those two, in the portion A of
the boundary, we extrapolate the toroidal component of the magnetic
field from the computational box to the boundary.

Such a choice of boundary conditions still leaves us with over-specified
boundary conditions, but this was the best combination we could obtain
through the whole parameter space. For some values of $\eta$, simulations
could be performed with the boundary conditions chosen closer to strict
mathematical demands, but not for all the resistivities presented
here. For a study of the dependence of the solutions on large resistivity, we
needed a set of simulations in which we could investigate the same solution
with an increasing parameter $\eta$.

In our computations we used various resolutions and sizes for the
computational domain. Here we present the results for a resolution
$R\times Z=(128\times 256)$ grid cells $=([0,50]\times [6,106])R_0$, in
the uniform grid. Results comply with the solutions for one quarter, one
half and double of this resolution, which we also computed for
verification. In our results we show only the $([0,50]\times [6,106])R_0$
part of the domain, but computations were performed with a three times
longer domain in the radial direction, $R\times Z=(384\times 256)$ grid cells
$=([0,150]\times [6,106])R_0$. The reason is that we preferred not to
additionally specify the outer boundary to avoid artificial
collimation as described in \citet{ust99}. We avoided doing so because in
the case of large resistivity we do not have any clue about the solution.
Instead, we extended the computational domain threefold in the radial
direction. We take the
result before any bounced wave from the outer-R boundary would reach a
part of the domain inside the $=([0,150]\times [6,106])R_0$. 
It turned out that in simulations with very large resistivity the
magnetic field does not collimate, and artificial collimation is
not an issue.

Here we give a summary of our {\sc pluto} code setup. We used cylindrical
coordinates, an ideal equation of state, and the ``dimensional splitting''
option, which uses the Strang operator splitting to solve the equations in
multiple dimensions. We checked if this introduces any difference, and found
that in our problem the results are not affected by this choice. The spatial
order of integration was set to ``LINEAR'', meaning that a piecewise TVD
linear interpolation is applied, accurate to second order in space.
We used the second order in time Runge Kutta evolution scheme RK2,
and for constraining the $\nabla\cdot\vec{B}=0$ at the truncation level,
we chose the Eight-Waves option. Instead of a Riemann solver, we used a
Lax-Friedrich scheme (``tvdlf'' solver option in {\sc pluto}).

\subsection{Conserved integrals}
\begin{figure*}
\hspace*{-1cm}\includegraphics[width=4.5cm,height=6cm]{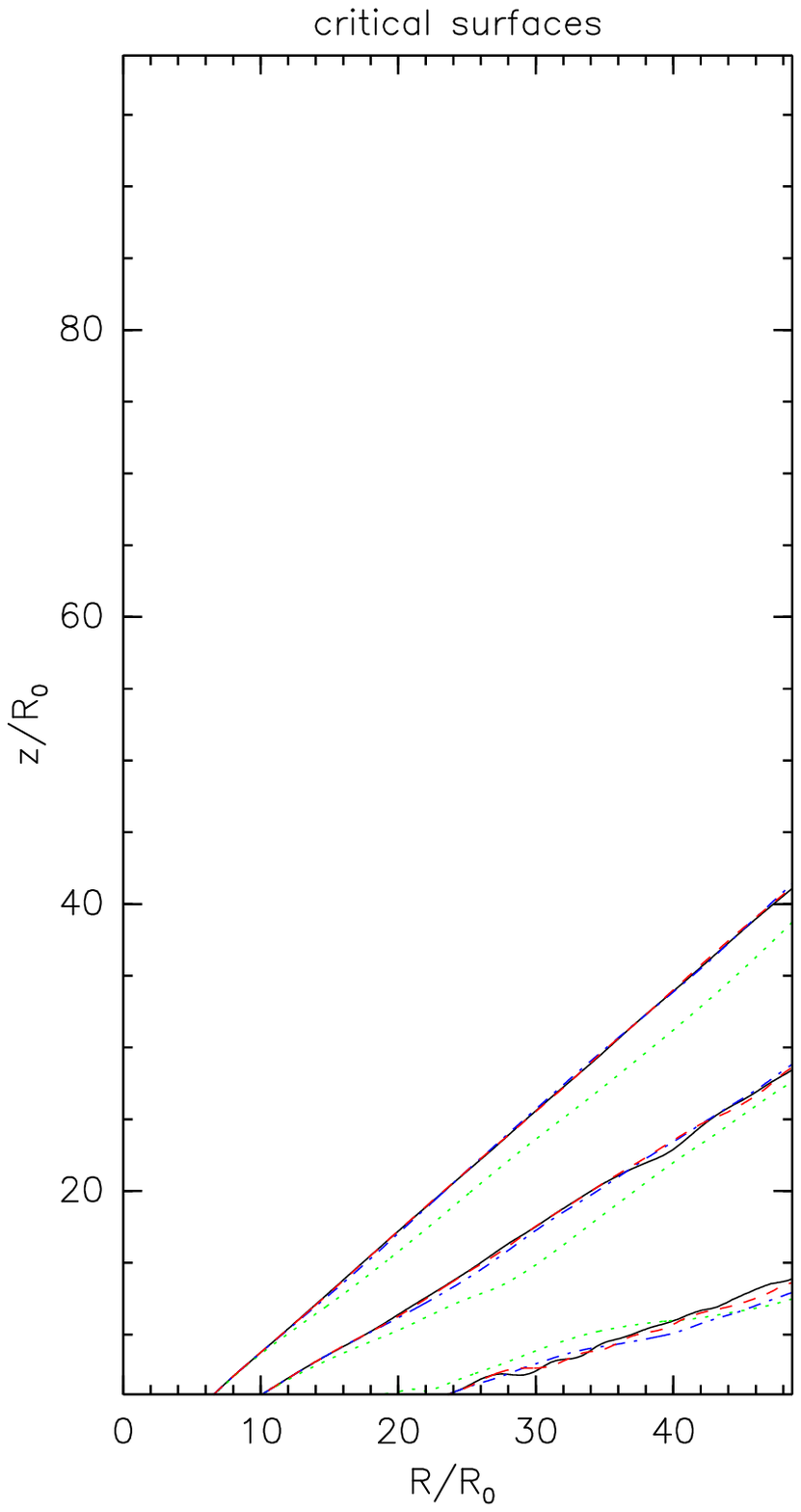}
\hspace*{0cm}\includegraphics[width=4.5cm,height=6cm]{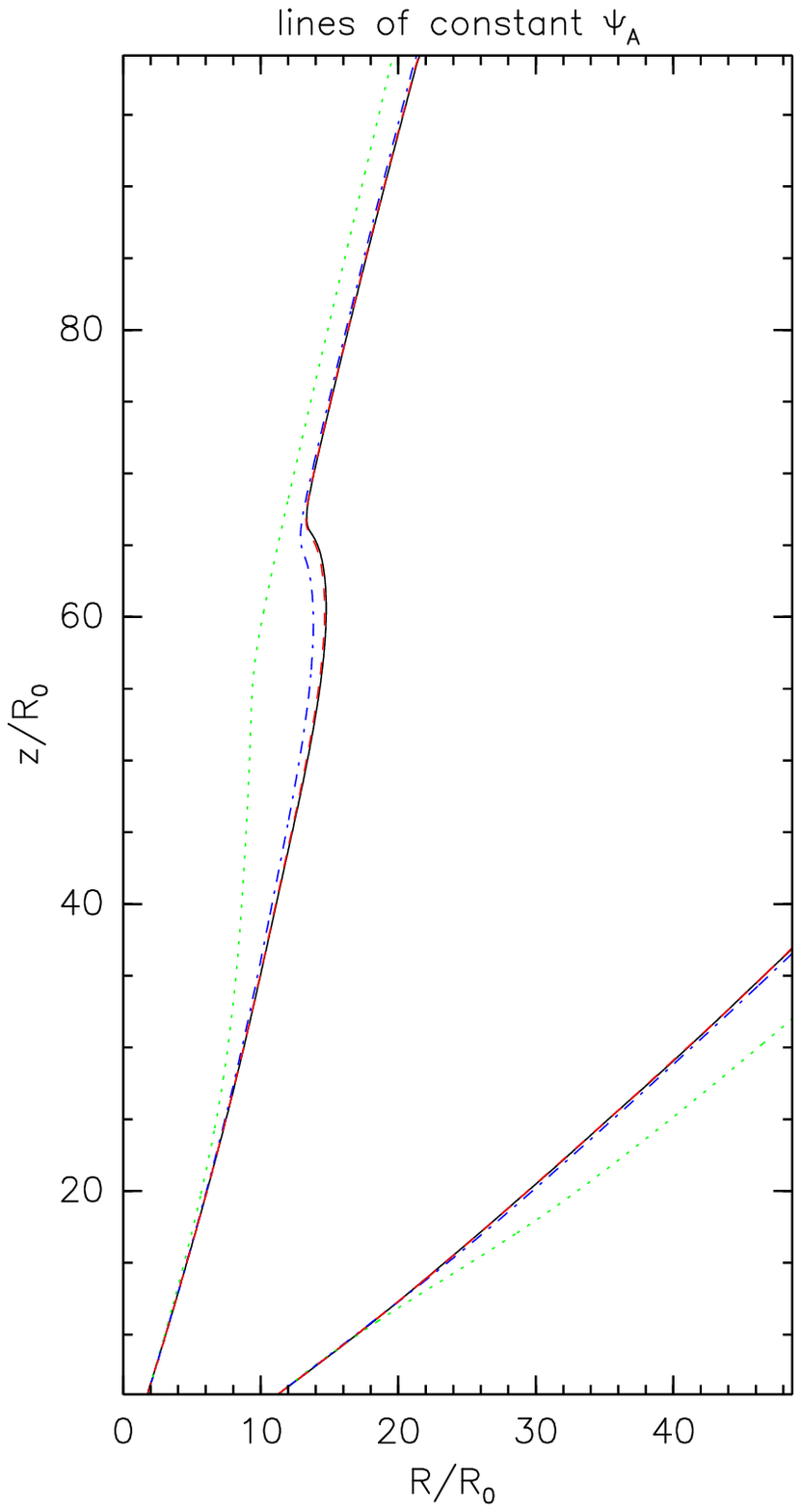}
\hspace*{0cm}\includegraphics[width=4.5cm,height=6cm]{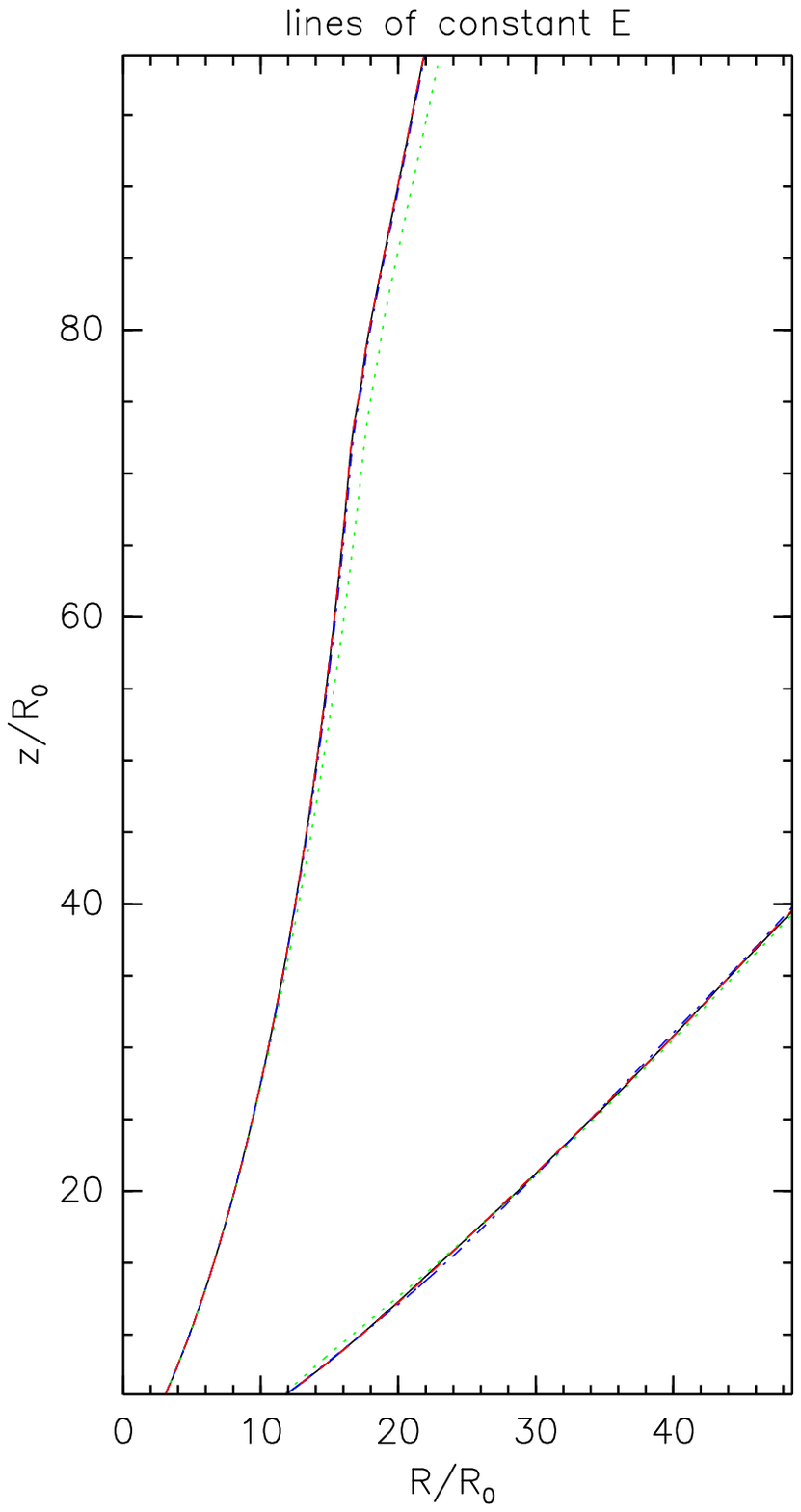}
\hspace*{4.5cm}\includegraphics[height=5.5cm]{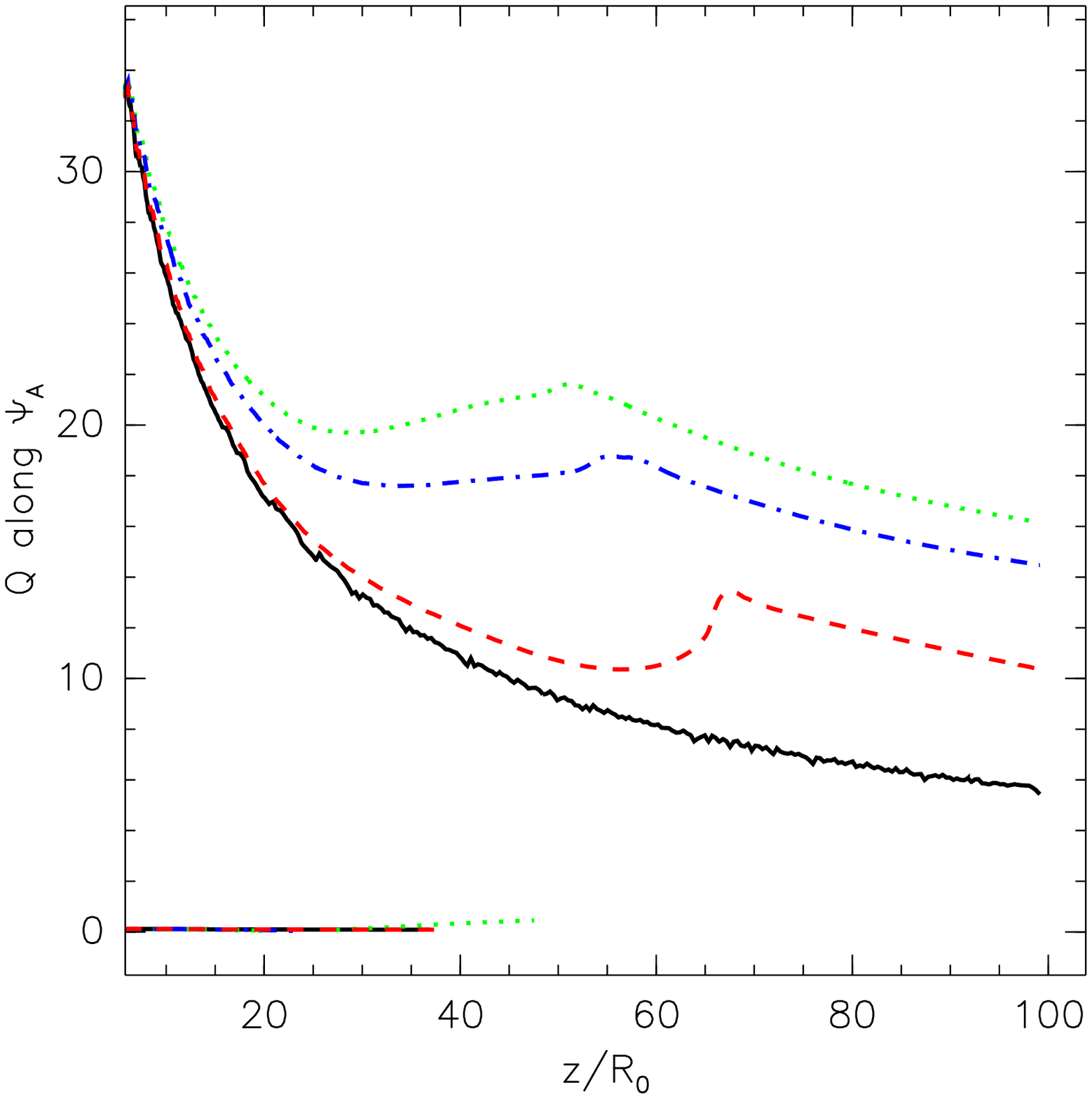}
\includegraphics[height=5.5cm]{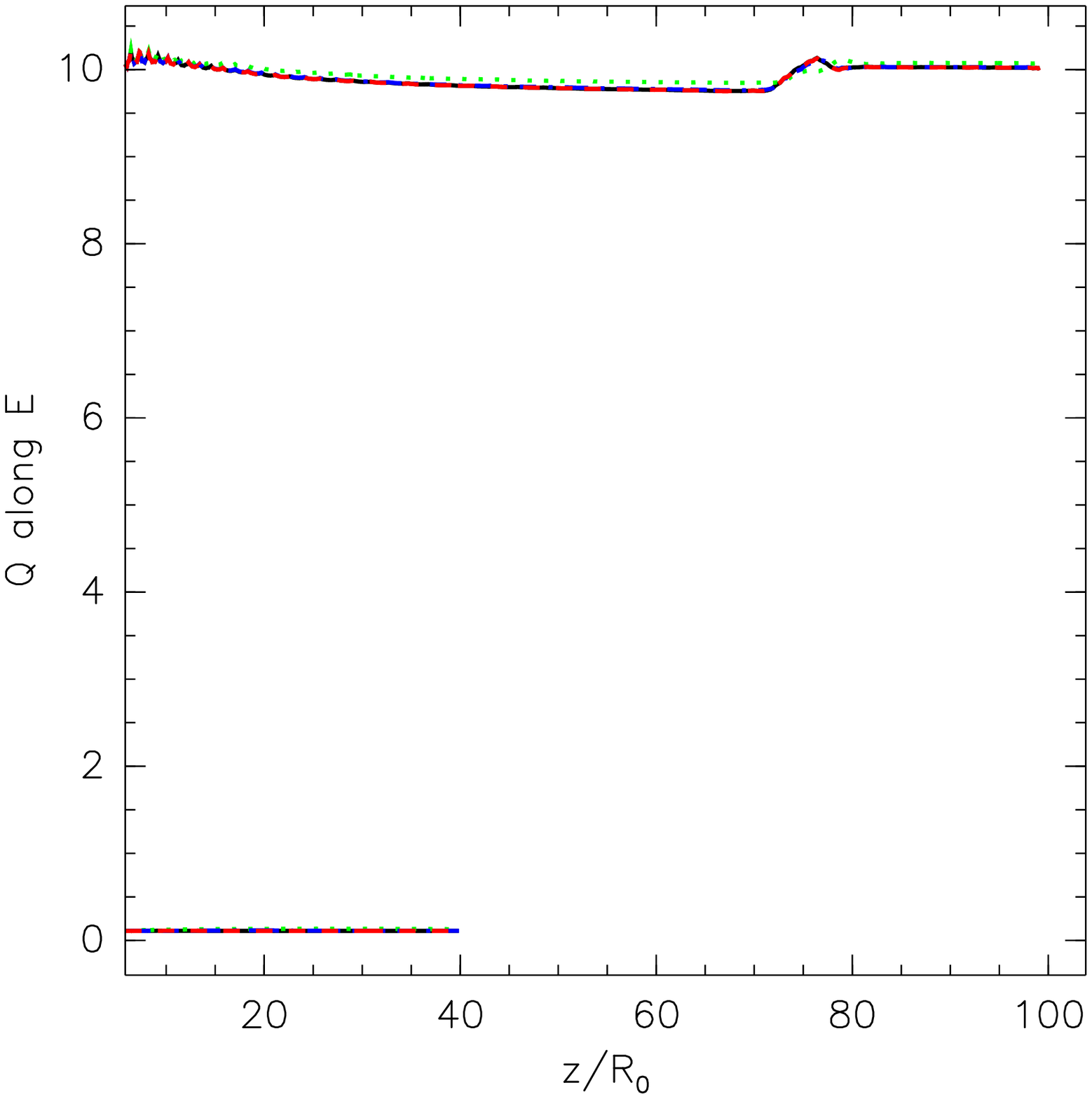}
\hspace*{4.5cm}\includegraphics[height=5.5cm]{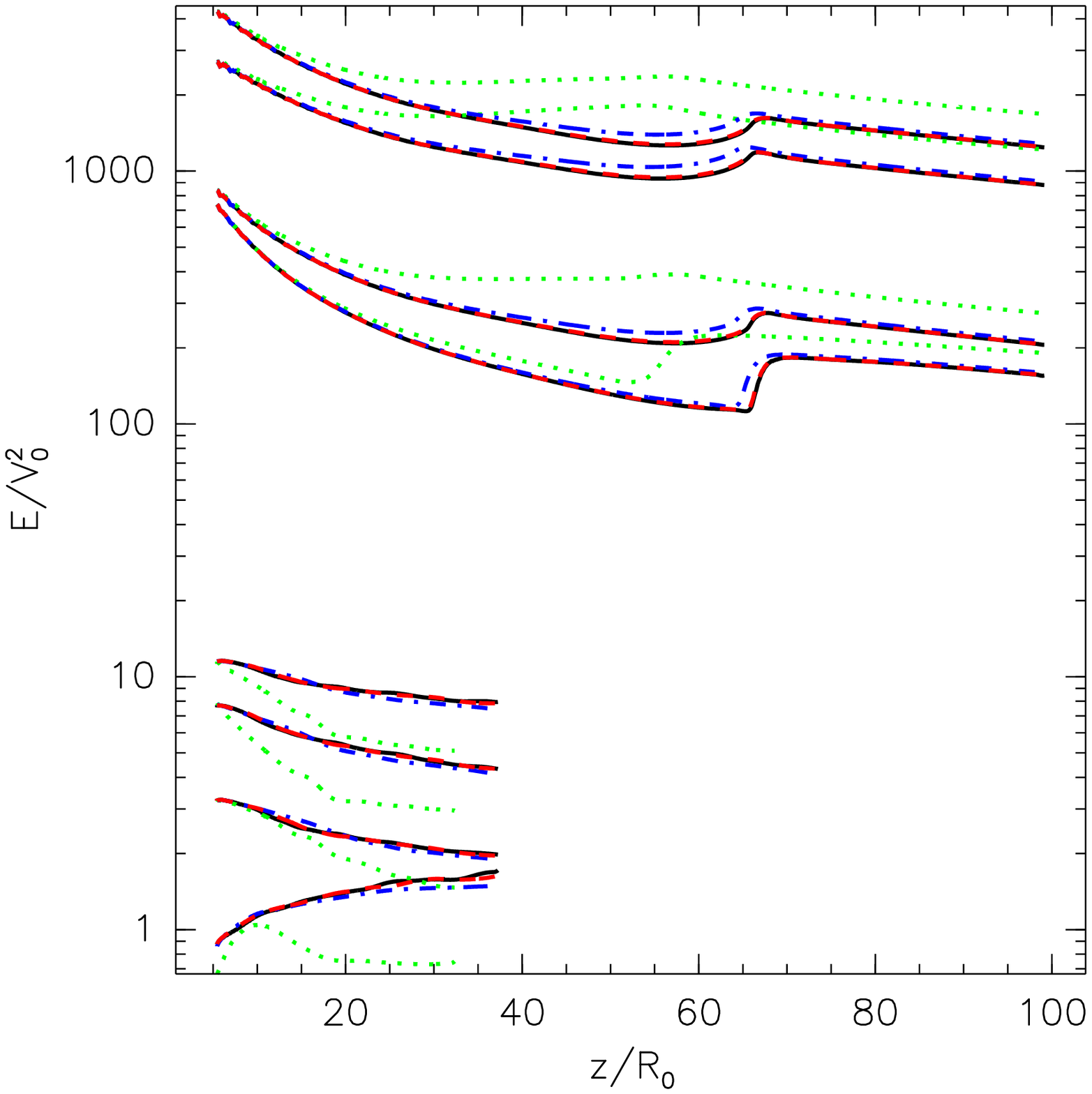}
\includegraphics[height=5.5cm]{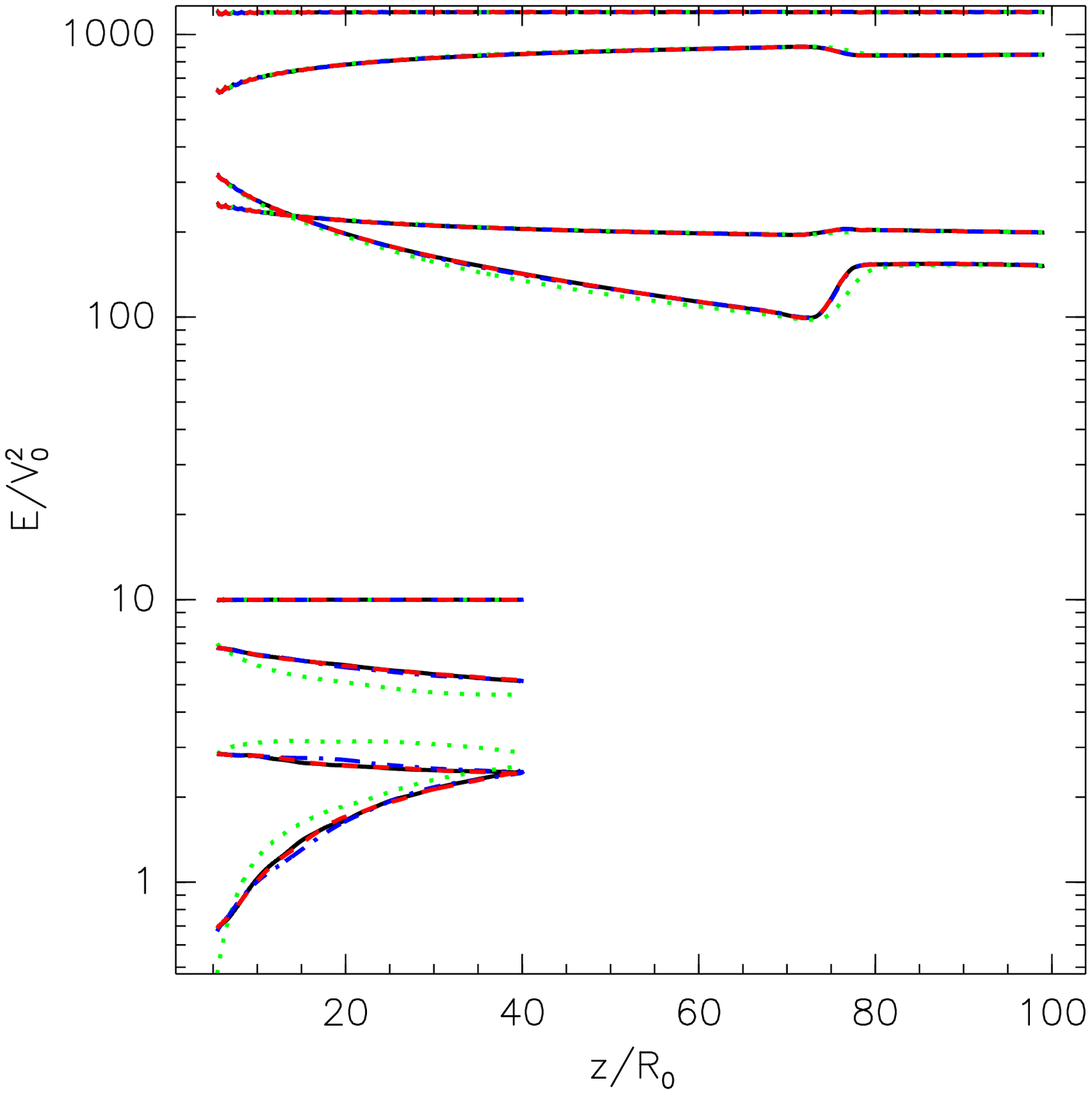}
\caption{Illustration of the effect of small physical
resistivity on the alignment of MHD integrals and magnetic flux
surfaces. In the {\em Top Left} panel we show critical surfaces in the
outflow with the resistivities $\eta=0,0.01,1$ and 1.5 shown in solid
(black), dashed (red), dot-dashed (blue) and dotted (green) lines,
respectively. The last value is shown to illustrate the difference in
comparison with solutions above the critical value of
$\eta_{\mathrm c}=1.0$. The Fast magnetosonic, Alfv\'{e}n and slow
magnetosonic critical surfaces are positioned from higher to
lower positions in the box, respectively. In the {\em Top Middle}
and {\em Top Right} panels are shown the shapes of two different
poloidal magnetic field lines and energy integral isocontour lines
along which we compute quantities shown in the panels in the rows
below (shown with the same line type and color as in the plot of
critical surfaces). As an example of the MHD-integrals,
in the {\em Middle} panels we show the entropy Q along those lines,
normalized to its value at large distance. In the {\em Bottom} panels
we show a split-down of the energy contributions along the same two  
lines. In both panels the upper set of curves corresponds to the
inner flux line, and the lower set of curves to the outer flux line.
The color code of the lines is for the resistivities as above. The
different line types now represent energy $E$ ({\em solid}),
kinetic energy ({\em dotted}), enthalpy ({\em dot-dashed}) and
Poynting flux energy ({\em dashed}). The gravitational energy is not
shown as it is orders of magnitude smaller.
}
\label{smallEtalines}
\end{figure*}
\begin{figure*}
\includegraphics[width=4.5cm,height=6cm]{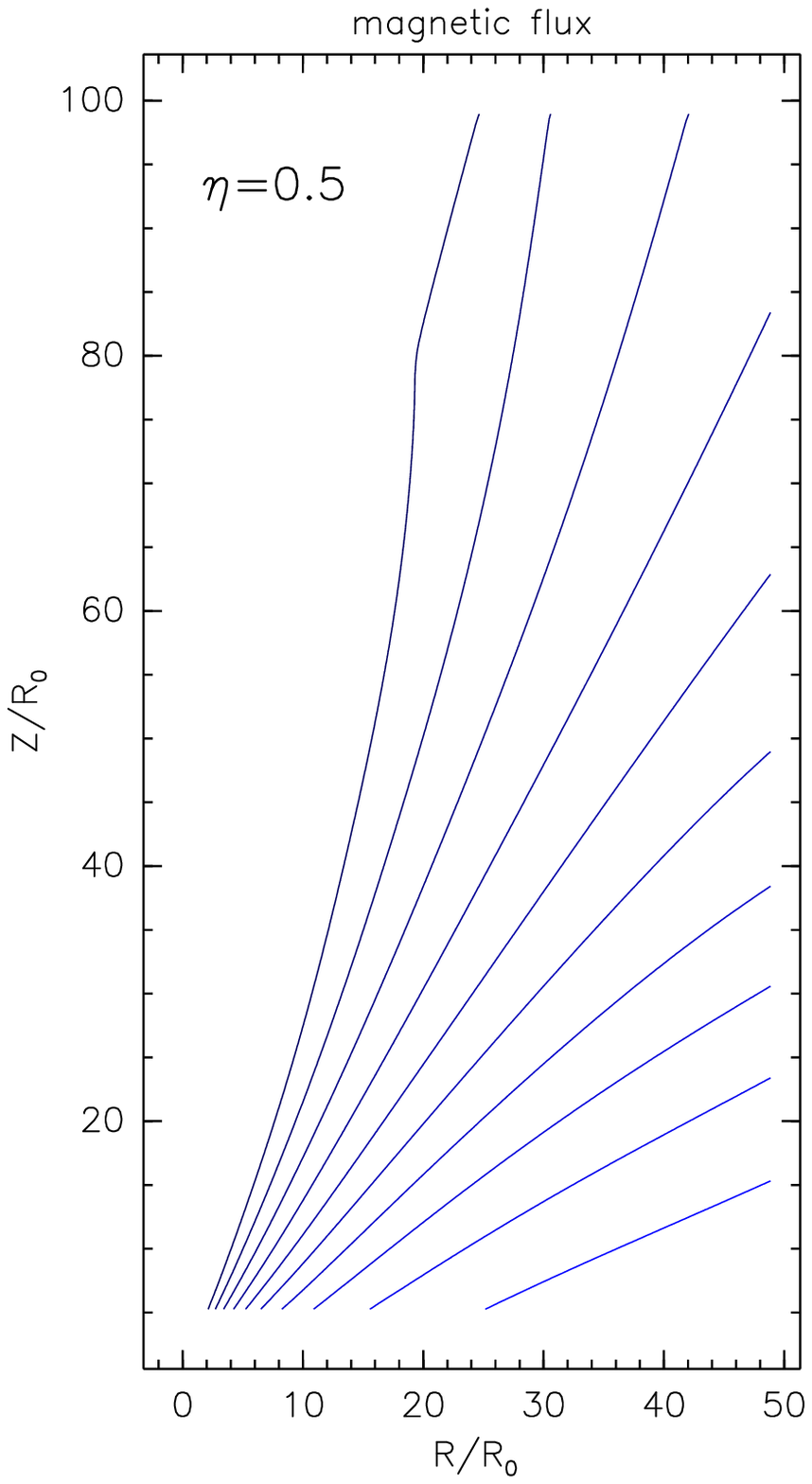}
\includegraphics[width=4.5cm,height=6cm]{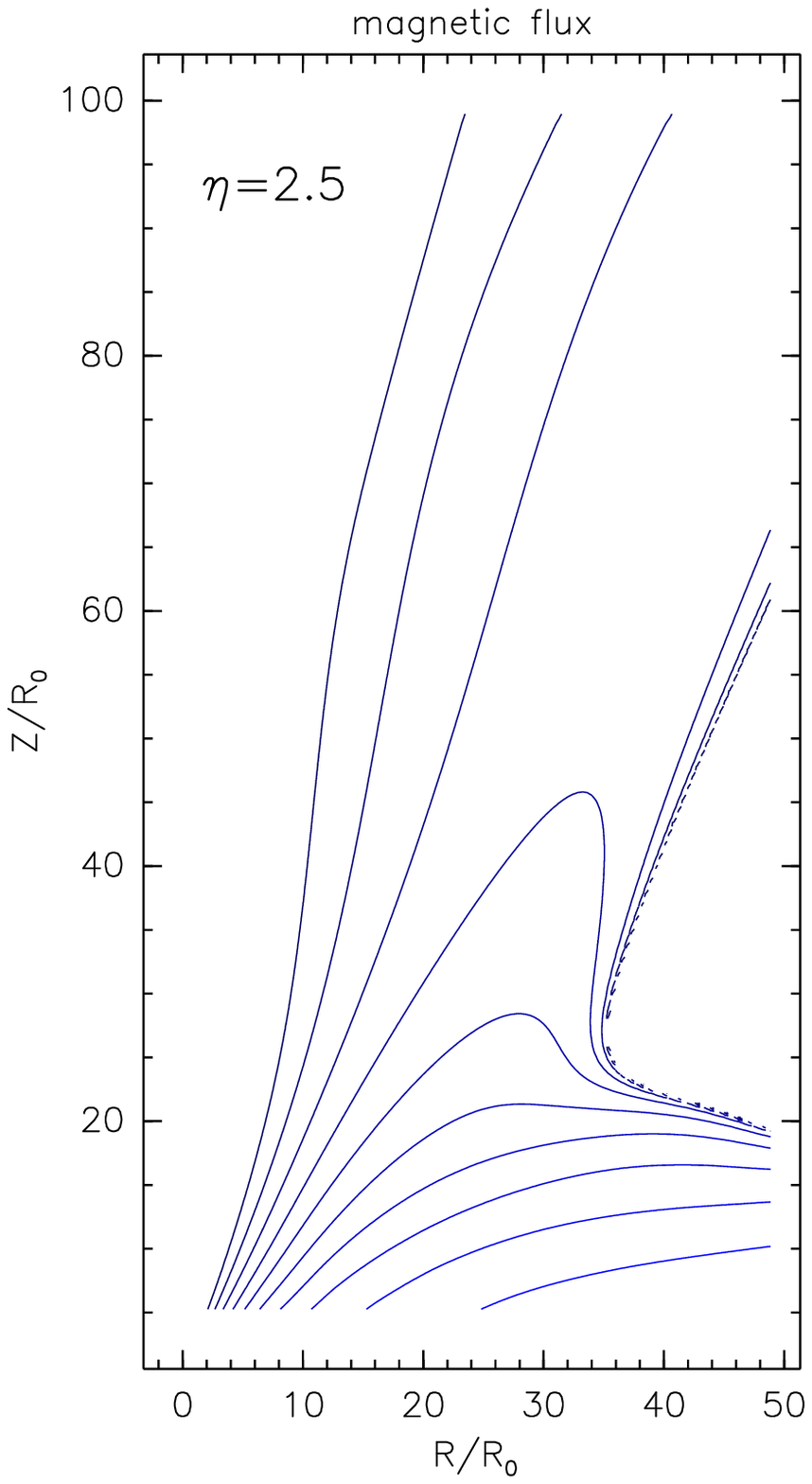}
\includegraphics[width=4.5cm,height=6cm]{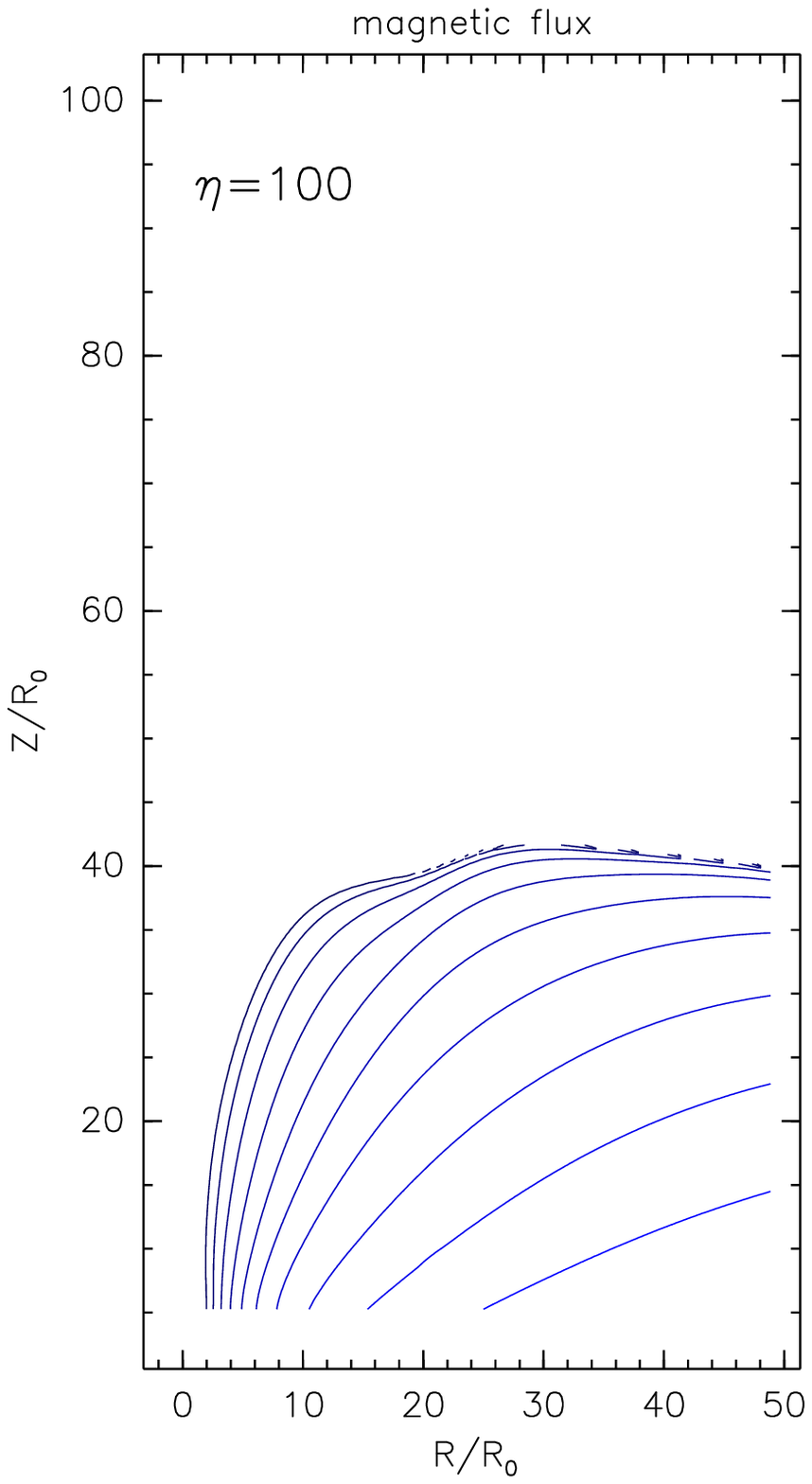}
\includegraphics[width=4.5cm,height=6cm]{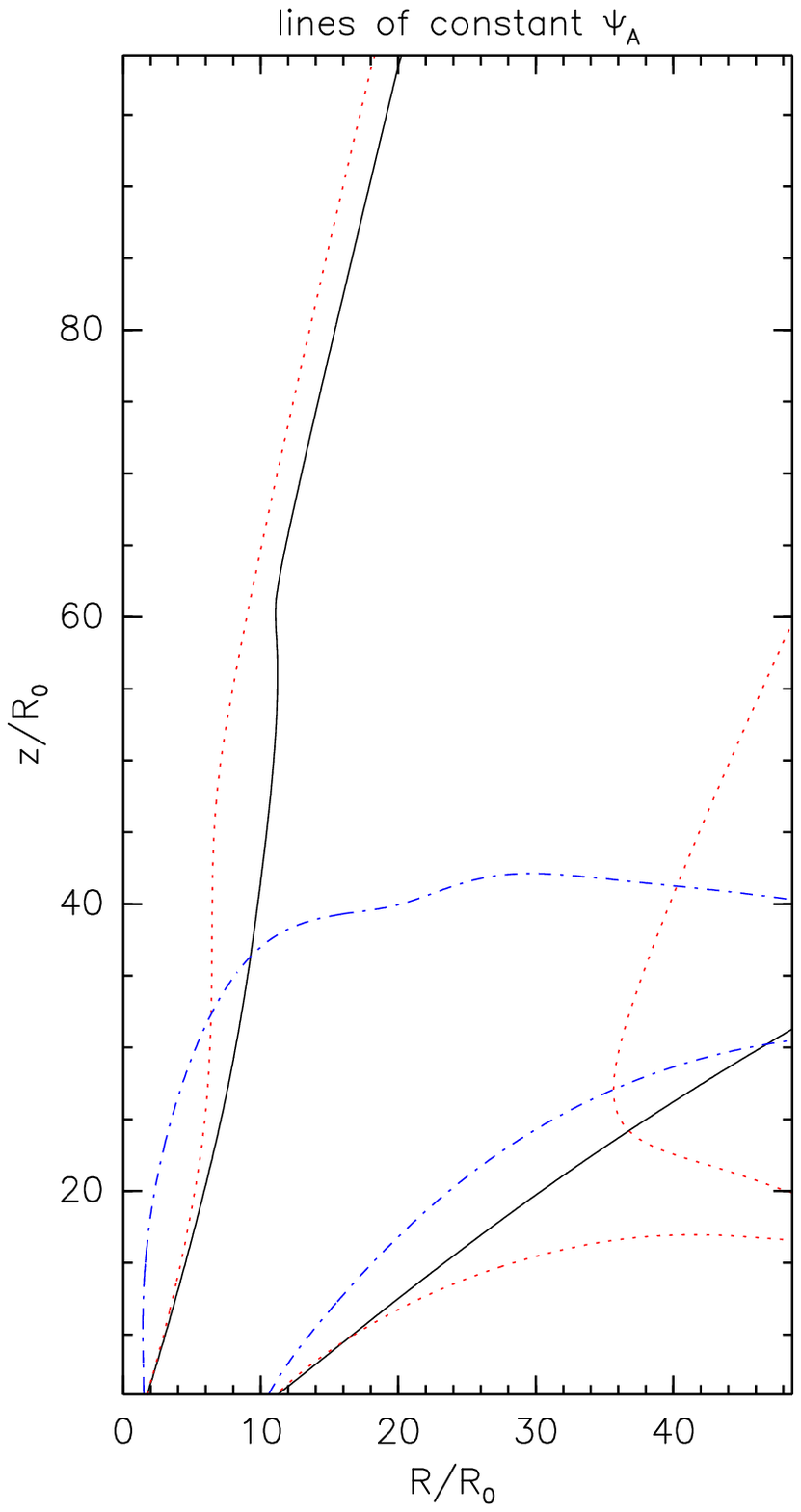}
\includegraphics[width=5.5cm]{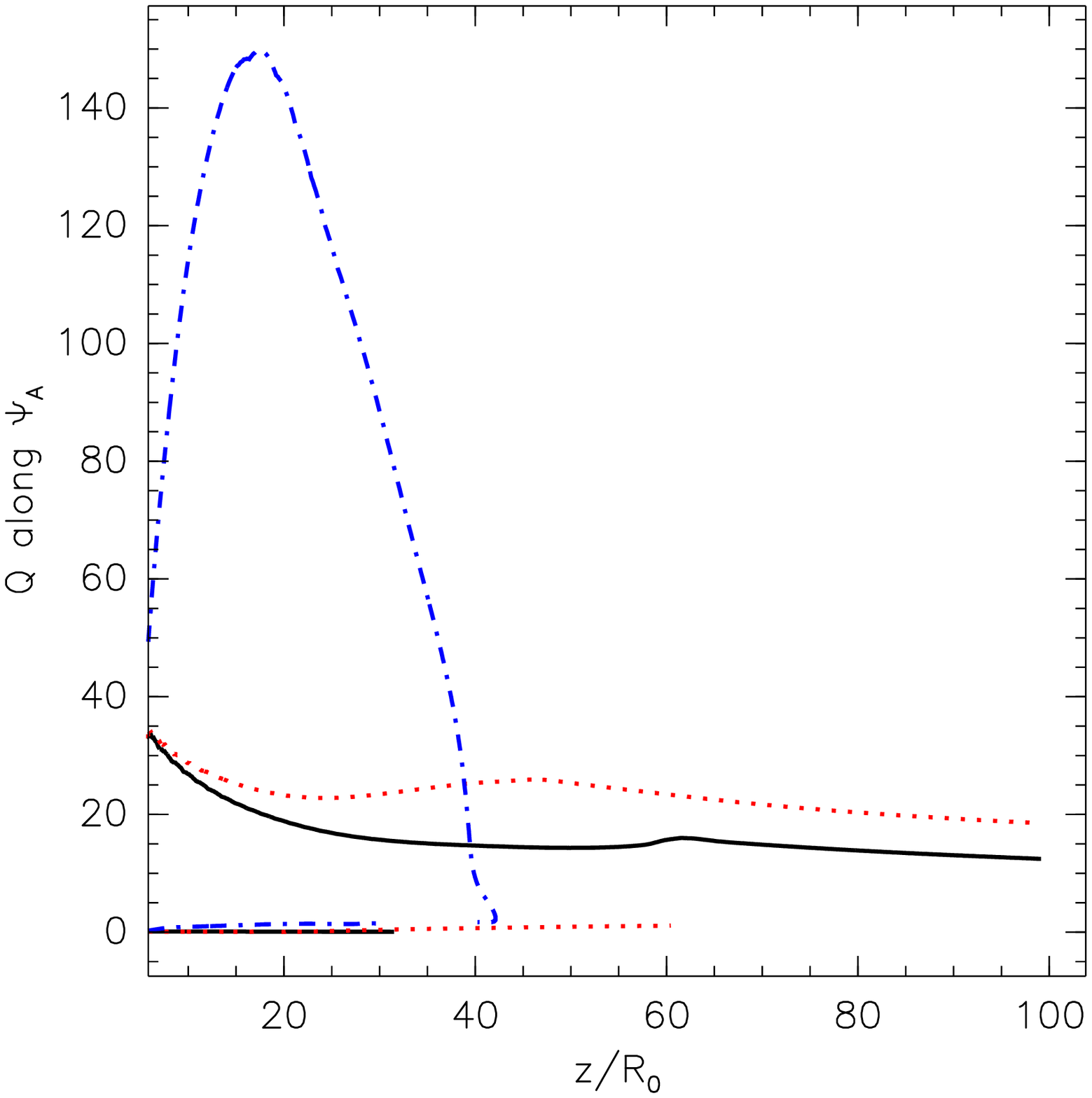}
\includegraphics[width=5.5cm]{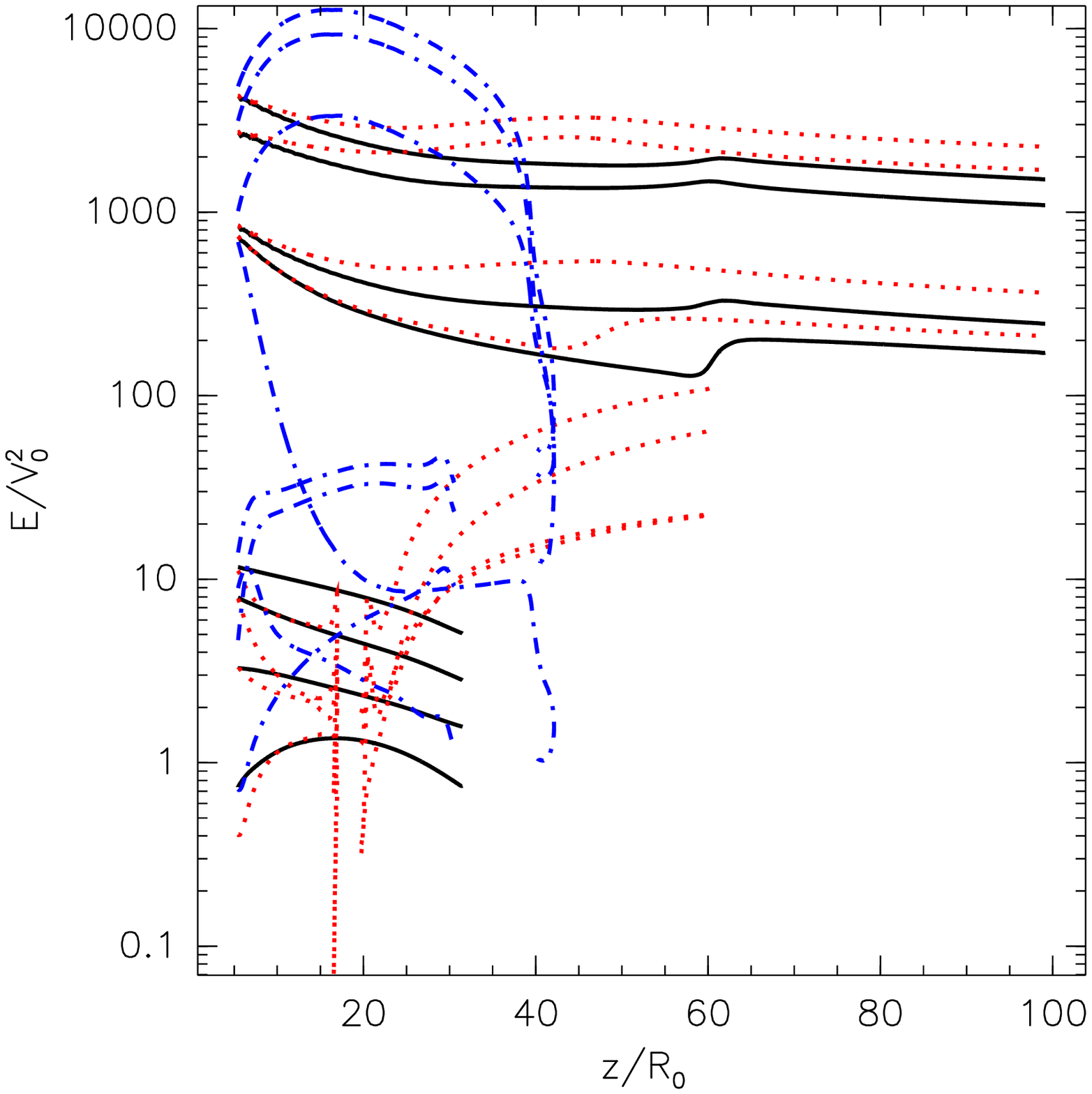}
\includegraphics[width=4.5cm,height=6cm]{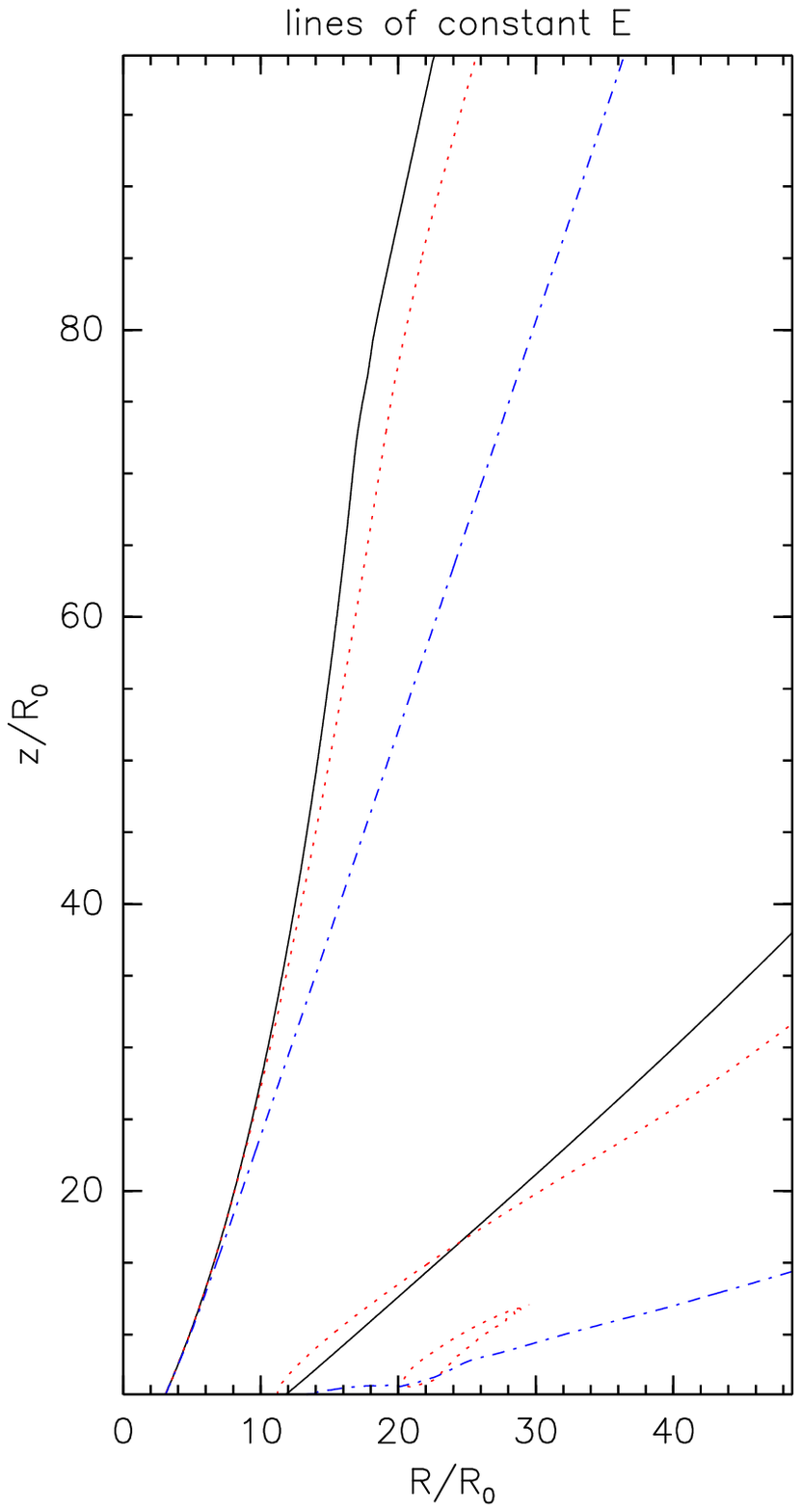}
\includegraphics[width=5.5cm]{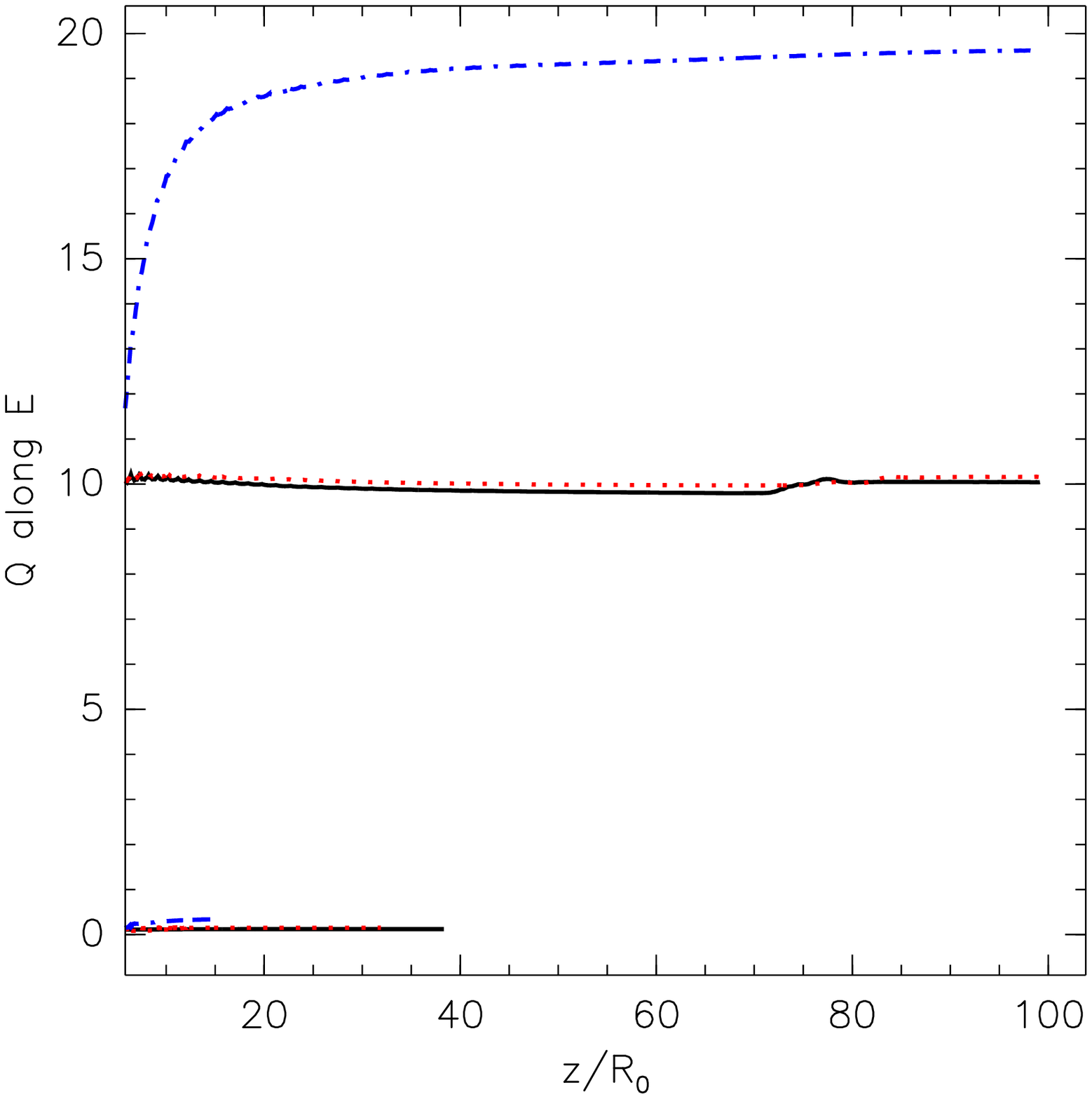}
\includegraphics[width=5.5cm]{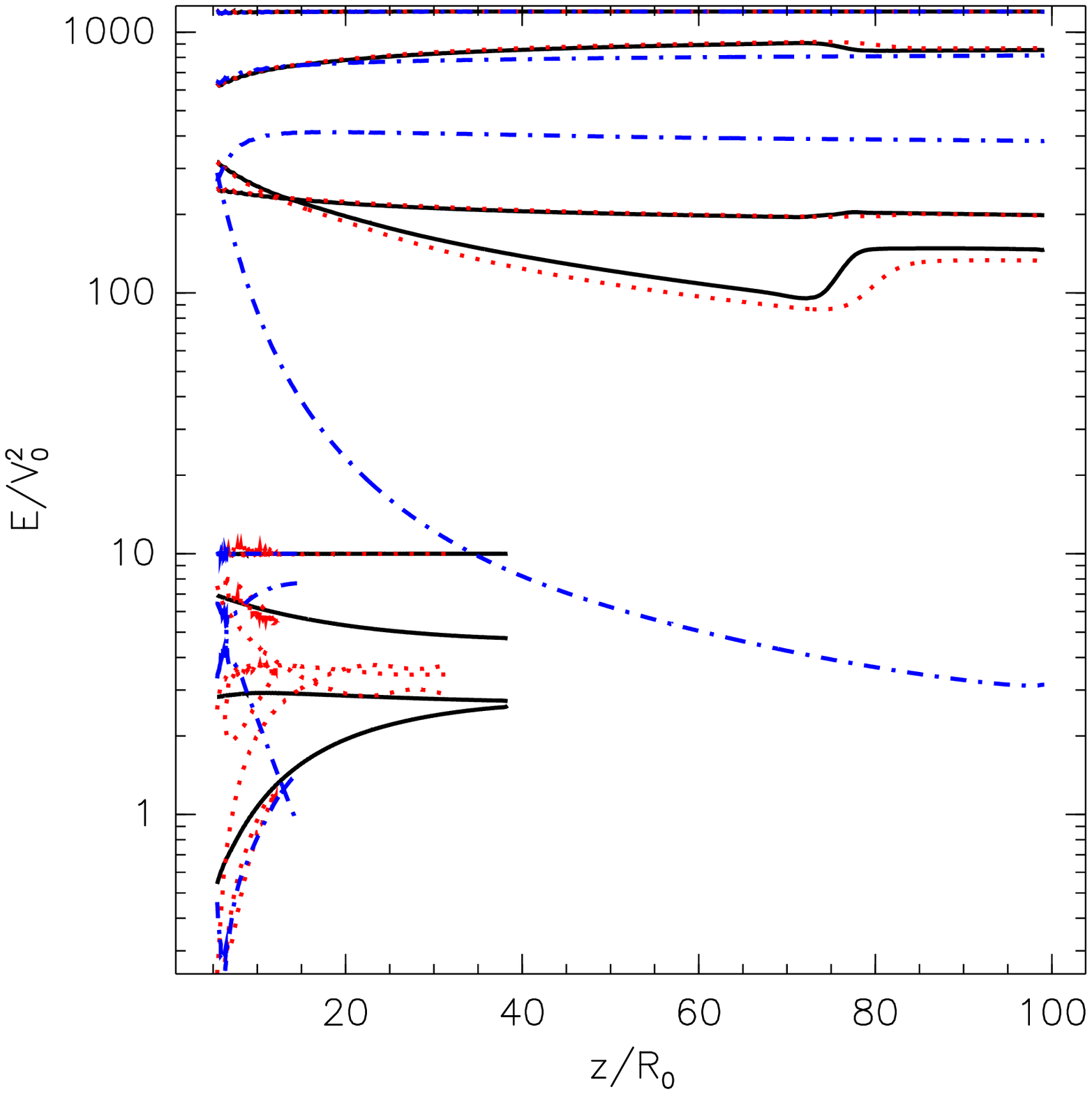}
\caption{Illustration of the effect of small, large and very large
physical resistivity on the magnetic field
lines and alignment of MHD integrals and magnetic flux surfaces.
In the {\em Top} panels are shown shapes of magnetic flux
isocontour lines (which are parallel to the poloidal magnetic
field lines) in a quasi-stationary state in simulations with
$\eta=0.5,2.5,100$ in the {\em Left}, {\em Middle} and
{\em Right} panels, respectively. In the
{\em Middle} and {\em Bottom} panels, the {\em Left}
panels show different poloidal magnetic field and energy
integral isocontour lines along which quantities in the panels
to their right are computed, for each resistivity. Legend of the
figures is the same as in Figure \ref{smallEtalines}.
}
\label{largeEtalines}
\end{figure*}
\begin{figure*}
\includegraphics[width=4.cm,height=6cm]{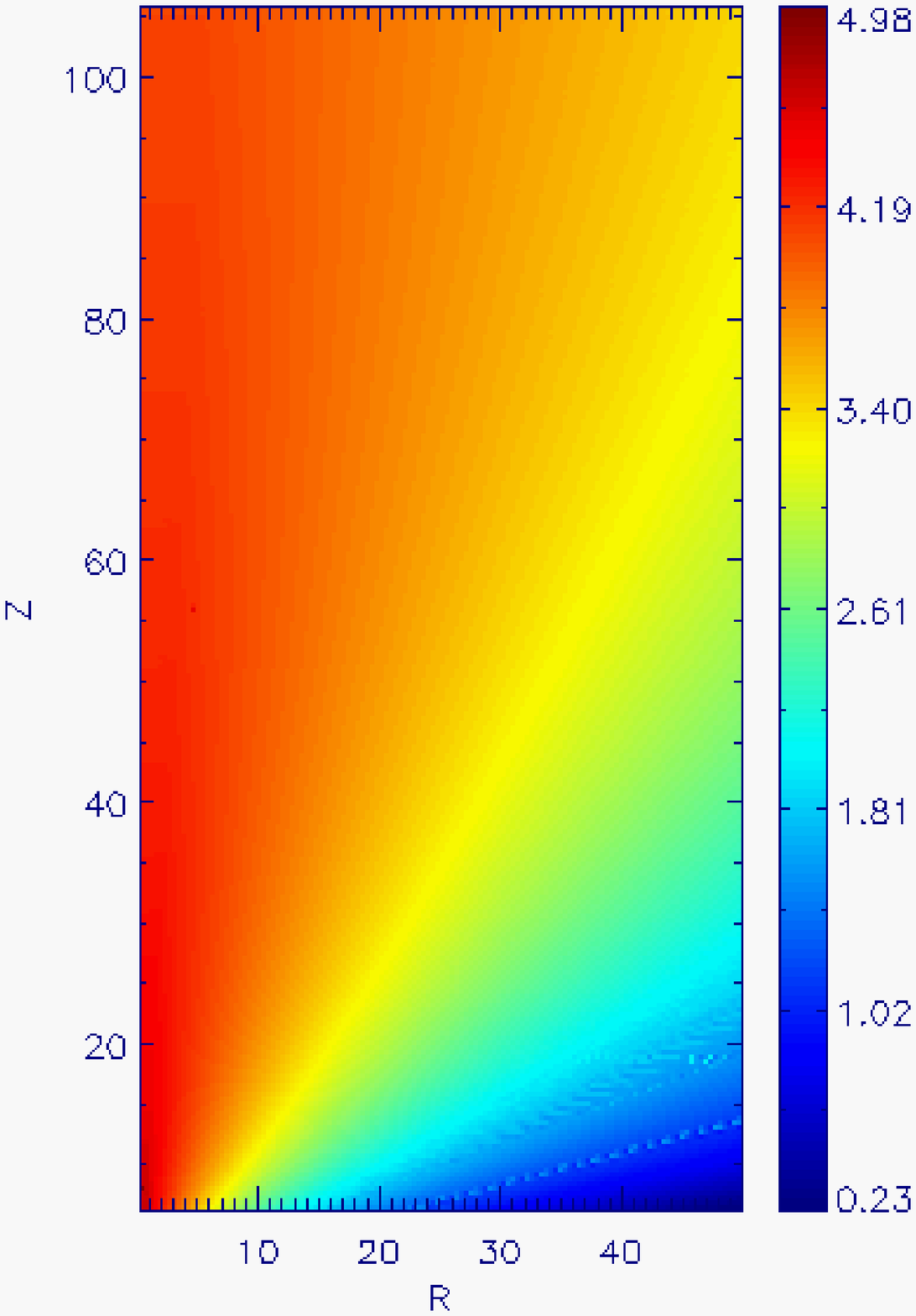}
\includegraphics[width=4.cm,height=6cm]{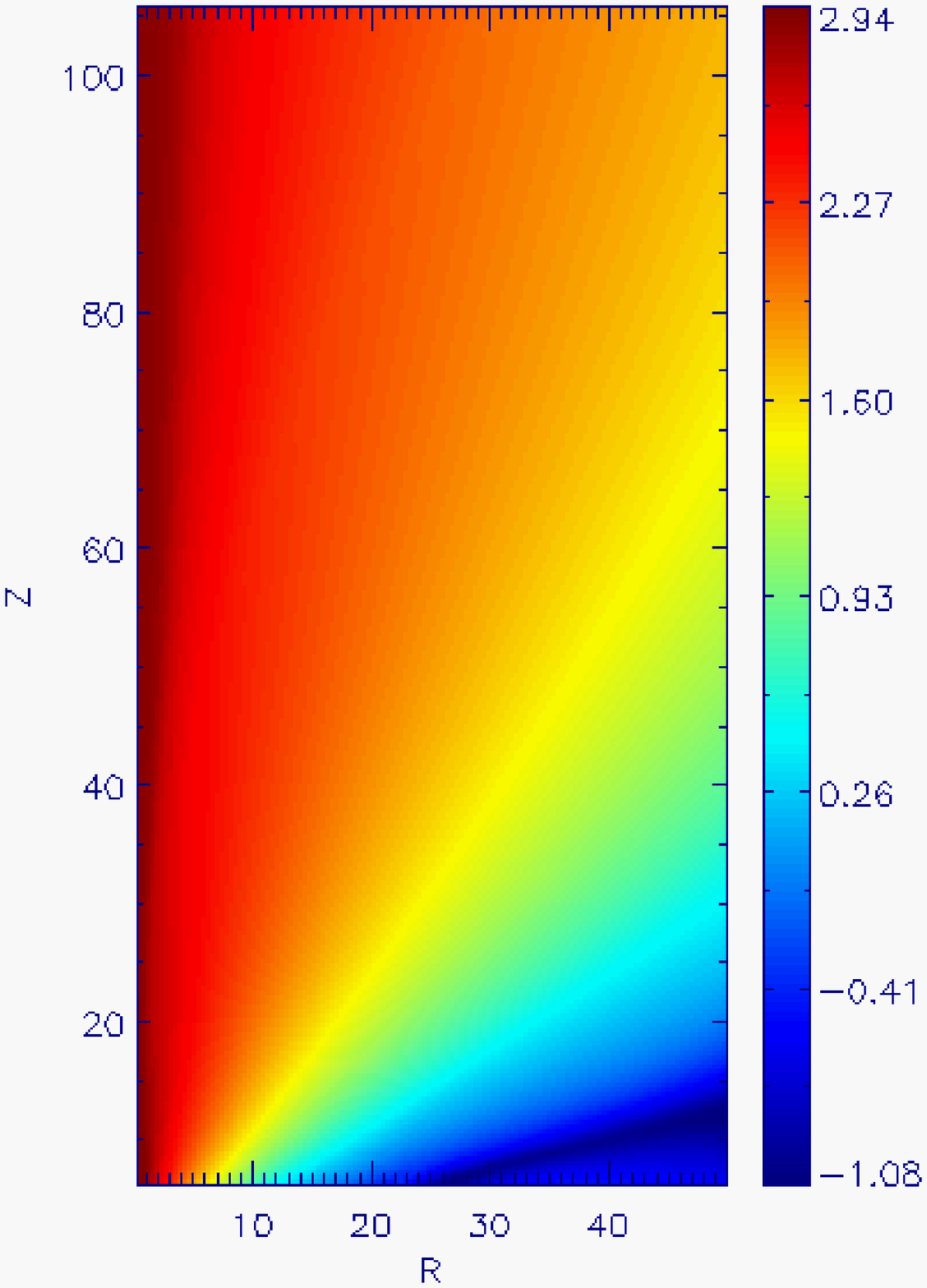}
\includegraphics[width=4.cm,height=6cm]{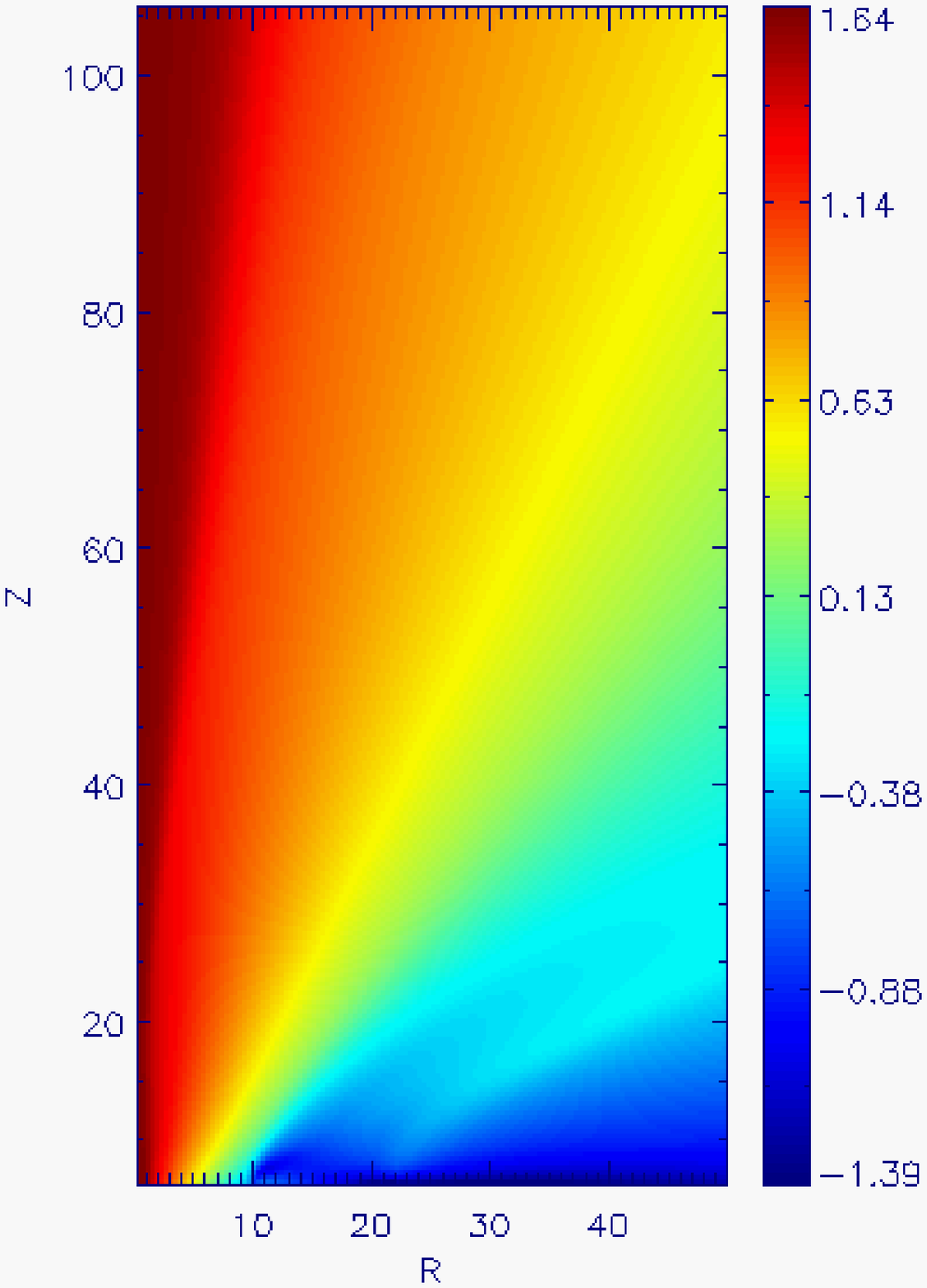}
\includegraphics[width=4.cm,height=6cm]{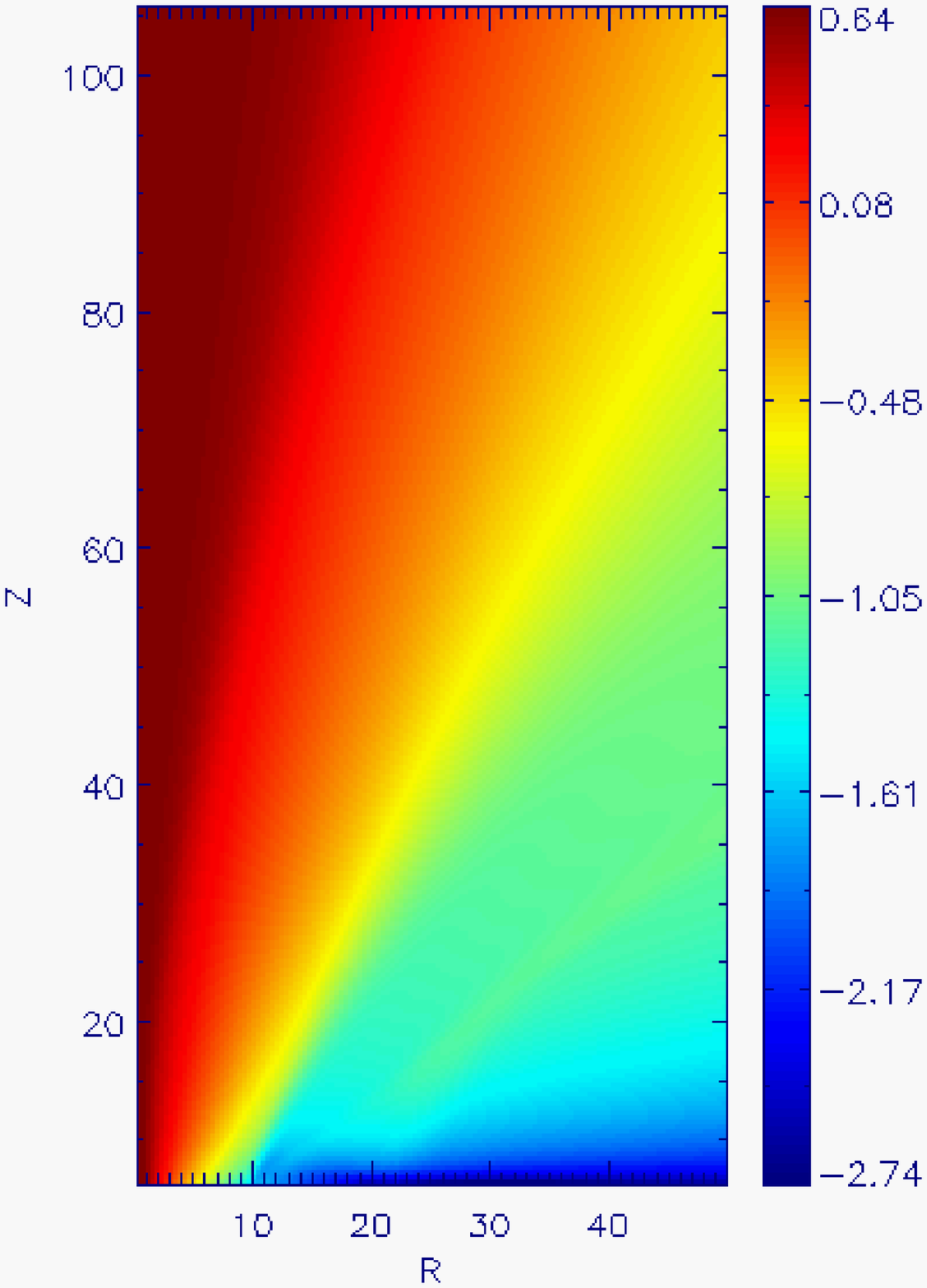}
\caption{Reynolds magnetic number Rm$=V\Delta x/\eta$ in the case of
initial conditions, small, large, and very large physical resistivity,
shown in the logarithmic color grading. {\em Left} to {\em Right} panels
show Rm for $\eta=0.01,0.5,10$ and 100, respectively. The numerical
resistivity, $\eta=0.01$ is plotted at T=0, for easier comparison of the
other values of $\eta$ to the solution from \citet{V00}.
}
\label{etas}
\end{figure*}
\begin{figure*}
\includegraphics[width=5.5cm,height=6cm]{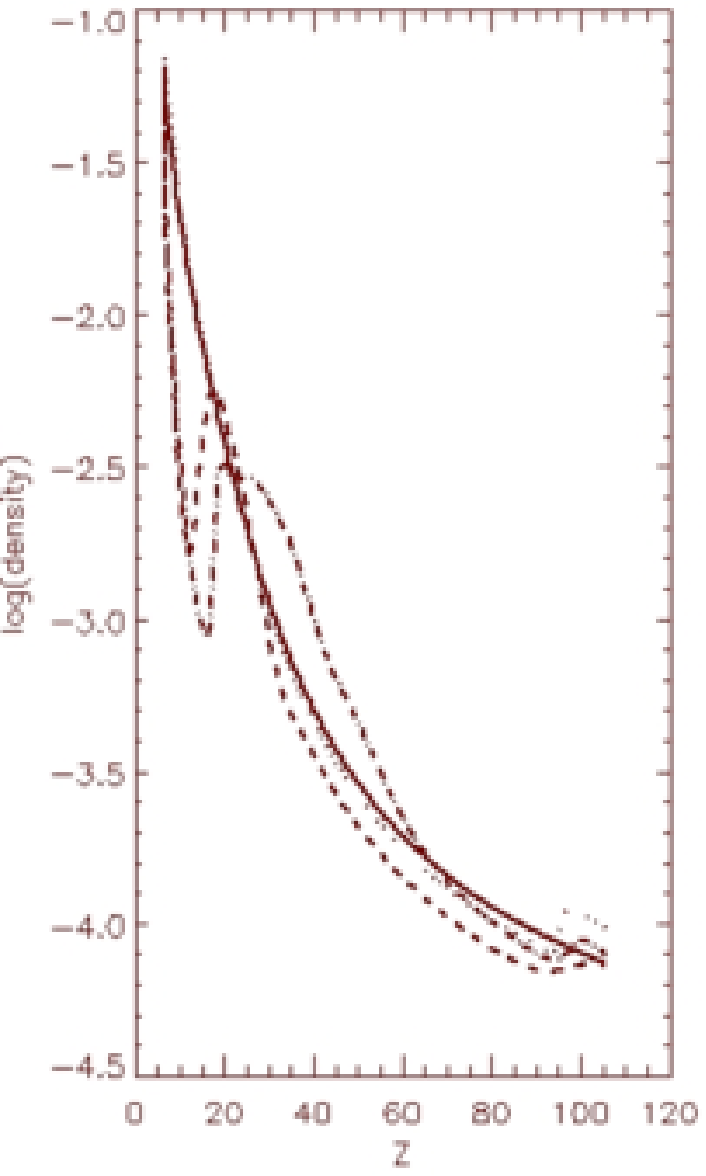}
\includegraphics[width=5.5cm,height=6cm]{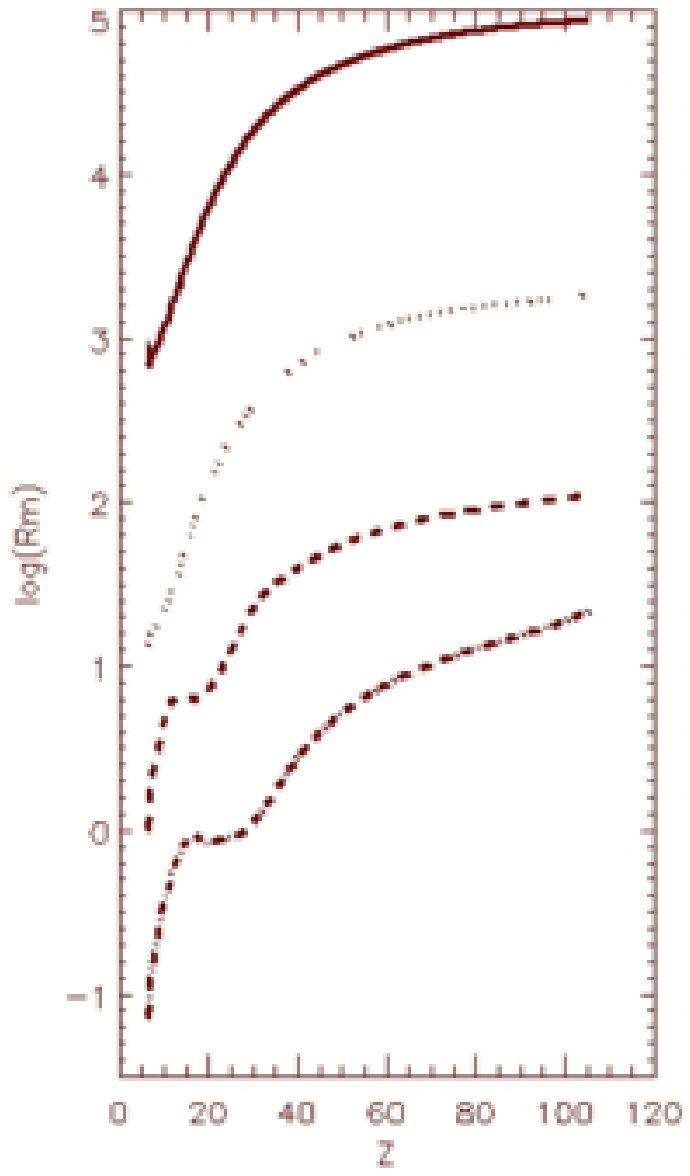}
\includegraphics[width=5.5cm,height=6cm]{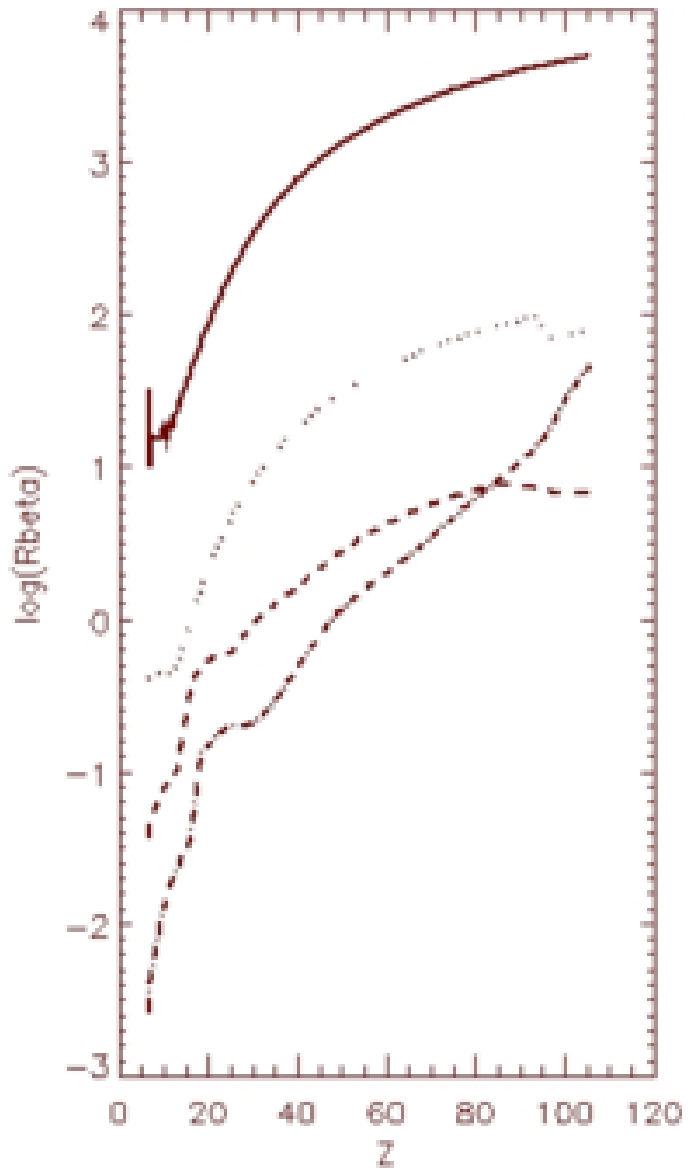}
\includegraphics[width=5.5cm,height=6cm]{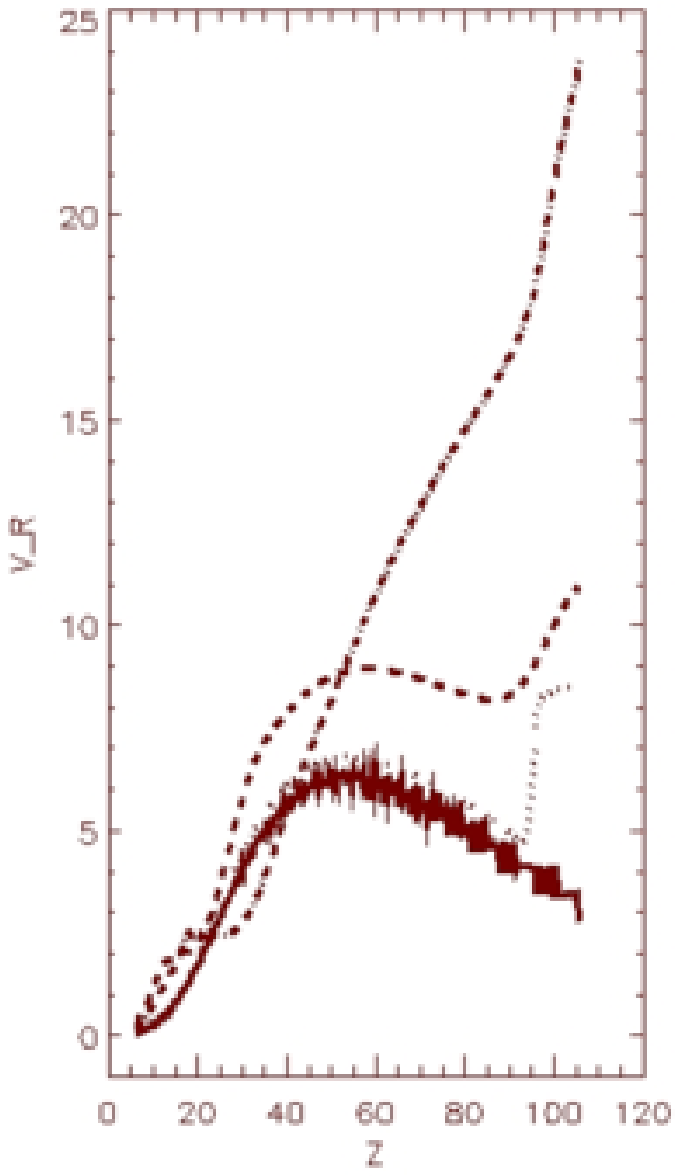}
\includegraphics[width=5.5cm,height=6cm]{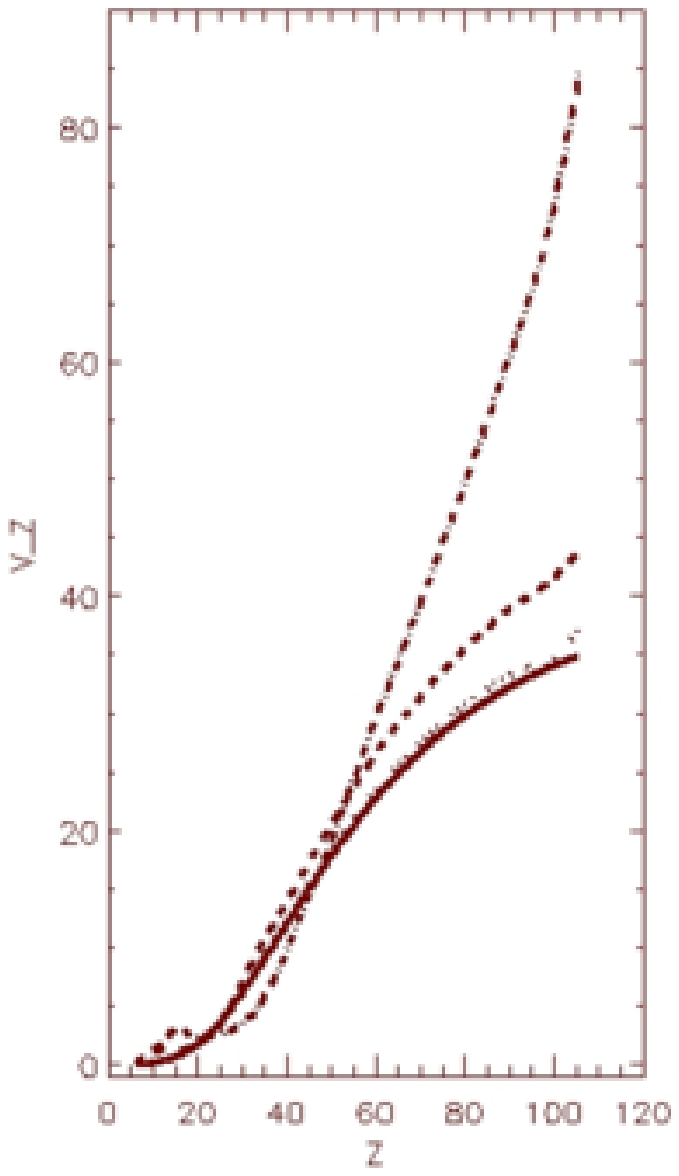}
\includegraphics[width=5.5cm,height=6cm]{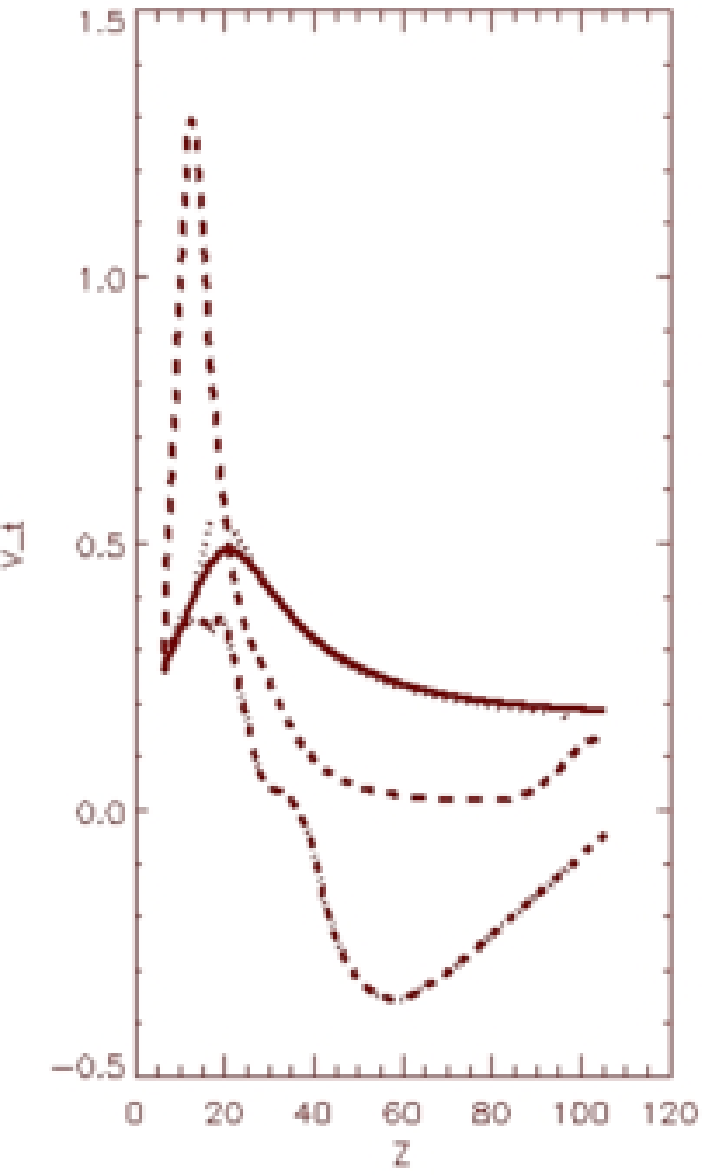}
\includegraphics[width=5.5cm,height=6cm]{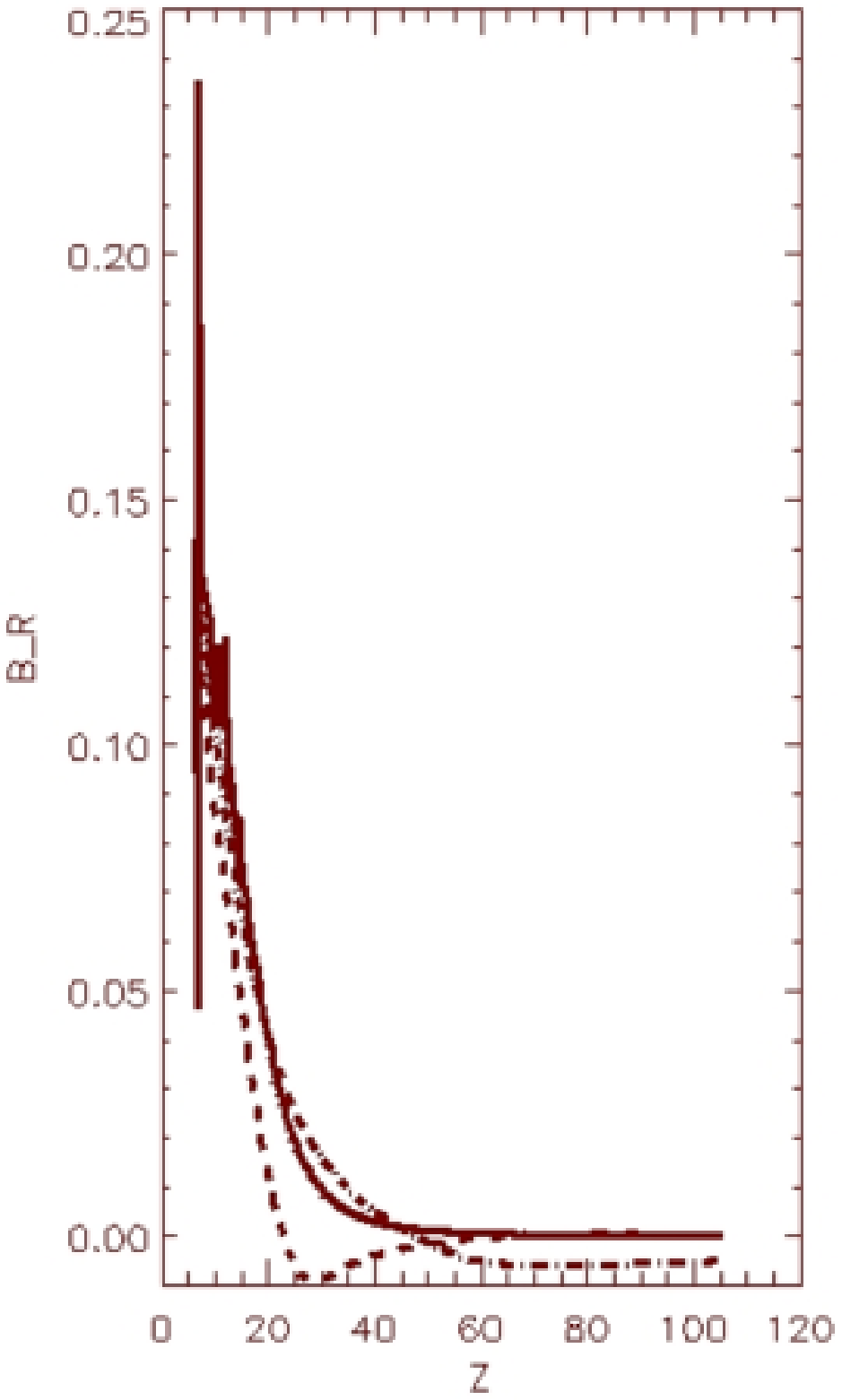}  
\includegraphics[width=5.5cm,height=6cm]{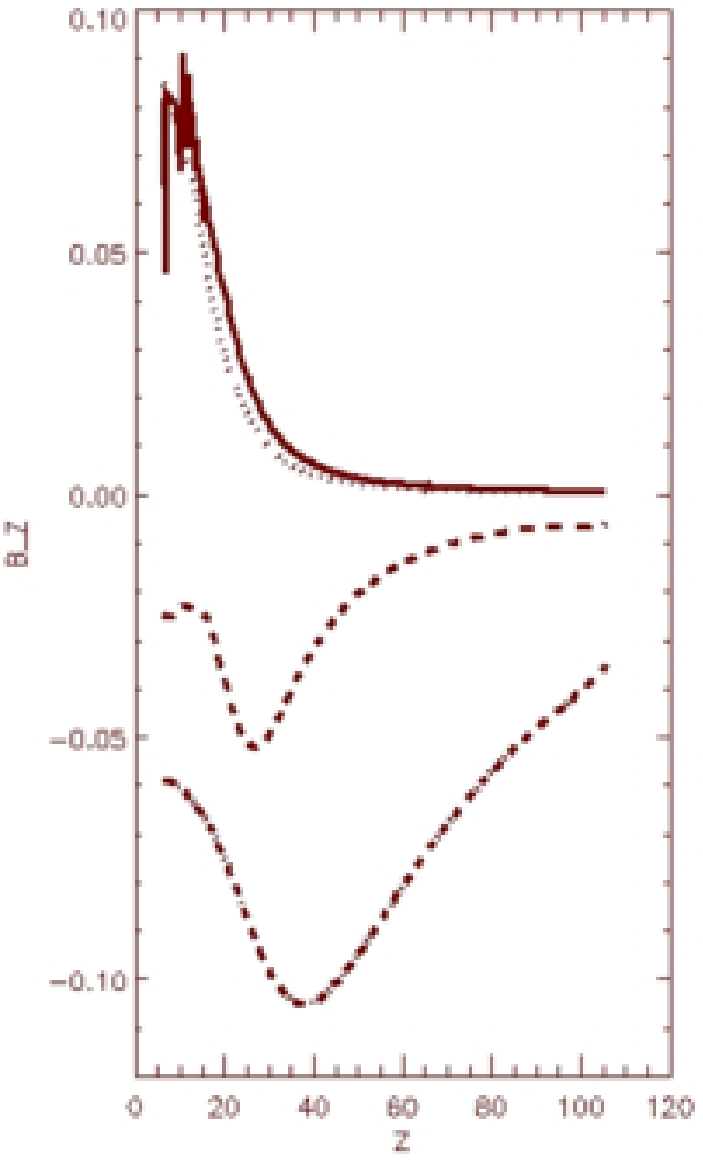}  
\includegraphics[width=5.5cm,height=6cm]{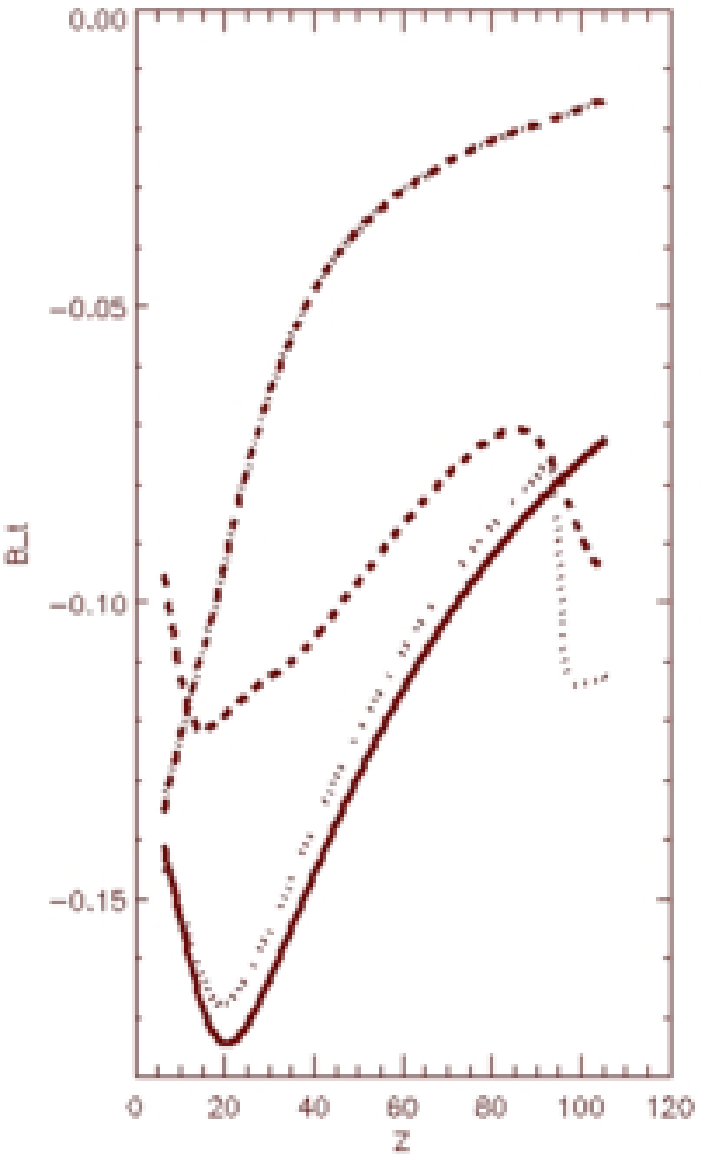}   
\caption{Various quantities computed along the slice parallel to the
symmetry axis, at half of the computational box (R=25).
In solid, dotted, dashed and dot-dashed lines for $\eta=0.01,0.5,10$
and 100 are shown density, Reynolds magnetic number and
R$_\beta=\mu_0 Rm P/B^2$ in logarithmic scale, and R, Z and toroidal 
components of velocity and magnetic field, respectively.
}
\label{varetasall}
\end{figure*}

It can be shown that steady, axisymmetric, ideal-MHD polytropic flows
conserve five physical quantities along the poloidal magnetic field lines
\citep{T82}. Those so called {\em integrals} are the
mass-to-magnetic-flux ratio $\Psi_A$, the field angular velocity $\Omega$,
the total angular momentum-to-mass flux ratio $L$, the entropy $Q$, and
the total energy-to-mass flux ratio $E$ (we call the latter
integral energy for brevity). They are given as
\beqa
  \Psi_A = \frac{4 \pi \rho V_p}{B_p} \,, \Omega=\frac{V_\phi}{R}-\frac{B_\phi}{B_p}
\frac{V_p}{R}\,,\\
  L = R V_\phi - \frac{R B_\phi B_p}{\mu_0 \rho V_p} \,, Q = p/\rho^\gamma \,, \\
  E = \frac{V^2}{2} + \frac{\gamma}{\gamma-1} \frac{P}{\rho}
  + \frac{ B_\phi \left( B_\phi V_p - B_p V_\phi \right)}{\mu_0 \rho V_p}
  -\frac{\cal GM}{r}
\,.
\eeqa
The various contributions of the energy $E$ correspond to the various
terms on the right hand-side of the equation. From left to right,
the kinetic, enthalpy, Poynting, and gravity terms
can be recognized.

The degree of alignment of the lines on the poloidal plane where the above
quantities are constant together with the poloidal magnetic field lines
can be used as a test how close to a steady-state the final result of
a simulation is.

\section{Numerical resistivity}
To investigate the effect of physical resistivity, it is necessary at
first to estimate the effect of {\em numerical} resistivity. We check
the level of numerical resistivity by comparing the simulations with
decreasing parameter $\eta$ until there is no effect on the solutions.
For simplicity, we choose the over-specified boundary conditions setup,
as was the case in C08. For small resistivities results are very
similar to the results with the proper number of boundary conditions
specified. Our results are presented in Figure \ref{numres}.

The numerical resistivity is estimated as $\eta=\Delta x^2/\tau$, where
$\Delta x$ and $\tau$ are the characteristic grid cell dimension and
the characteristic time scale of the physical process, in
this case diffusion of magnetic field. For the same $\tau$, the
difference in numerical resistivity is then dependent on the second
power of difference in grid cell dimensions. Explicitly, for grid
cells of dimensions $\Delta x_1$ and $\Delta x_2$,
$\eta_1/\eta_2=\Delta x_1^2/\Delta x_2^2$. This means that for
$\Delta x_2=0.5 \Delta x_1$ and $\eta_1/\eta_2=0.25$ the numerical
resistivity would be of the same order, and no clear difference could
be observed. In consequence, it is not enough to just double (or half)
the resolution, as is usually done, to check the effects of resistivity,
one has to go further. Only for $\Delta x_2=0.25 \Delta x_1$
do we obtain $\eta_1/\eta_2=0.06$, with the numerical resistivity now
different by an order of magnitude, the effect should be clearly visible.

One can not easily find a nontrivial problem with such a well defined
stationary solution to compare this relation for numerical
resistivities including all the terms in the MHD equations.
An estimate of numerical resistivity follows the theoretical prediction,
that its effects are visible only for a quadruple change in resolution.
Our problem is well suited for such a check. We verified this
prediction using our simulations in the low-resistivity range of
physical resistivity. The procedure we described here could be used
as a standard test for resistive MHD codes.

\section{Results with physical resistivity}
The case of small resistivity has been investigated in C08,
with the {\sc nirvana} code. With large resistivity, we obtained
a departure from the smooth pattern but, because of computational
restrictions, we could not address it. As we have changed the code for
our computations, we can now check the previously obtained results. We
investigate in detail what happens after the departure of the
solution from the smooth pattern of change in the position of critical
surfaces and integrals of motion.

In the simulations presented here, the computational box extends
three times further in radial direction than the length of the
portion of the box which we analyze. Therefore, we can control, and be
certain that any information eventually traveling back towards the
origin from the outer-R boundary did not reach the portion of the
domain in which we are interested. All the simulations reach the
quasi-stationary state before the disturbance from the outer boundary
reaches back into the R=[0,50] part of the box.
 
\subsection{Results with small resistivity}
As the first step in our simulations, we check if the setup in the
{\sc pluto} code gives the same trend observed previously in C08, with the
smoothly increasing departure of integrals of motion from the ideal-MHD
case for increasing resistivity. In Figure \ref{smallEtalines} we show
the positions of the critical surfaces for the diffusivity $\eta=$0, 0.01,
1.0 and 1.5. The critical $\eta$ we identify as $\eta_{\mathrm c}=1$.
From the results obtained we conclude that the trend observed in previous
simulations by the {\sc nirvana} code holds also in the {\sc pluto} results.
With increasing resistivity, the change in the integrals of motion and
the positions of critical surfaces are uniform and small. When reaching some
critical resistivity, $\eta_{\mathrm c}=1.0$, the solutions still largely
resemble the initial setup in the geometry of the magnetic field, but
the critical surfaces and integrals of motion step out of the smooth
pattern seen for the smaller values of $\eta$. An illustration of this
is shown in Figure \ref{smallEtalines}. In the bottom panels of Figure
\ref{densres} is shown the transition in geometry between the solutions
for small and large resistivity. This is an essential piece of
information, since results like these trends of change are directly
comparable only inside the same geometry. For the very different
geometries we should compare only more general, non-local trends.

\section{Results with large and very large resistivity}
With resistivities approaching $\eta=1$, the geometry of the solutions for
the magnetic field lines starts to change. The value of $\eta=1$ is a
critical value for this change, as it is where the smooth trend
described in the previous subsection no longer applies in the whole
box, because of the changed geometry. The typical solution
in this second mode of geometry for a radially self-similar setup is
shown in the middle panels of Figure \ref{densres}. The magnetic field
``bulges'' in the super-fast magnetosonic portion of the flow, and is
compressed towards the meridional plane at large radial distances.

With even larger resistivities, approaching $\eta=10$, the magnetic
field geometry drastically changes once again. The ``bulge''
acquired with large resistivities is smoothed out, and only the
compressed part of the field above the disk surface remains. This
third mode of geometry for a radially self-similar setup is shown in
the {\em Right} panels of Figure \ref{densres}. There is no longer any
collimation of the magnetic field. This mode of the solution is
stationary, it does not change further with an increase in resistivity.
The largest value of resistivity for which our code still could run to
a quasi-stationary state was of the order of a few hundreds.

Figure \ref{etas} shows the Reynolds magnetic number in our computational
box in simulations with increasing $\eta$. The profile of Rm
traces the increasingly de-collimating outflow, indicating that the resistivity
is directly related to the geometry of the magnetic field. In
Figure \ref{varetasall} we show various quantities along the slice
parallel to the axis of symmetry, at half the computational box, for
the full range of $\eta$ considered in this work. The density does not differ
much in our solutions. Poloidal velocity increases for large and very
large resistivities. Rotation changes direction for very large resistivities.
B$_Z$ also changes direction for large and very large resistivities, but
it is interesting that B$_R$ changes direction only for large, but not for
very large resistivities.

In \citet{C08} we estimated the physical magnetic diffusivity in the
astrophysical case of young stellar object with solar mass
${\cal M} \approx M_\odot$ and with a characteristic distance of
$0.1AU=1.4 \times 10^{10}$m. Here we extend it to the increased range of
diffusivity. From Eq.~(\ref{magdif}), for the diffusivities
$\etahat=(1,10,100)$ in the code units, we obtain the physical
diffusivities $\eta=(1,10,100)\times 6.8\times 10^{14}$m$^2$s$^{-1}=(1,10,100)\times
6.8\times 10^{18}$cm$^2$s$^{-1}$, respectively.

\section{Summary}
We obtained solutions of resistive-MHD self-similar
outflows above the accretion disk for the regimes of small, large and
very large resistivities. The disk is not included in the simulations, but is
taken as a boundary condition.

To find the lower limit of small resistivity, we first find the level of
numerical resistivity. We also verified the predicted behavior of numerical
resistivity with increasing grid resolution: it is not enough to
double the grid resolution to obtain the decrease in numerical resistivity
for an order of magnitude. One has to quadruple the number of grid cells.
A similar procedure could be used as a standard test for resistive MHD codes.

Next we verified the solutions from our previous work in C08, for small
resistivity, since there we used a different code. We find the same trend.
Since now we have a parallel code, we can investigate solutions with
larger resistivities. Solutions with small, large and very large
resistivities each have a different geometry, so that we can distinguish
three modes of solutions.

The first mode is similar to the ideal-MHD solutions. With $\eta>1$,
there appears a ``bulge'' in the magnetic field, which is a signature
of the second mode. At even larger resistivities, with $\eta>10$, the
third mode of solutions sets in. In the third mode, the magnetic field
is no longer collimated, but is pressed towards the disk. Such a solution
does not change further with increasing resistivity. 

\section{Acknowledgments}
M\v{C} completed this work in the Theoretical Institute for Advanced
Research in Astrophysics (TIARA) under contract funding by Theoretical
Institute for Advanced Research in Astrophysics (TIARA) in the Academia
Sinica and National Tsing Hua University through the Excellence Program of
the NSC, Taiwan. Simulations have been initially performed during
M\v{C} stay in Athens at spring 2009, supported by the European
Community's Marie Curie Actions - Human Resource and Mobility within the
JETSET (Jet Simulations, Experiments and Theory) network under contract
MRTN-CT-2004005592. Fruitful discussions with Ruben Krasnopolsky are 
acknowledged, and we thank A. Mignone for possibility to use the
{\sc pluto} code. M\v{C} thanks to M. Stute for help with {\sc pluto}
setup, and J. Gracia for his python scripts for some of the plots.

\bsp

\label{lastpage}

\end{document}